\documentclass[12pt]{article}
\usepackage[dvips]{epsfig}

\makeatletter
 
\def\@cite#1#2{{#1\if@tempswa , #2\fi}}
 
\def\section{\@startsection {section}{1}{\z@}{+3.0ex plus +1ex minus
  +.2ex}{2.3ex plus .2ex}{\normalsize\bf}}
\def\subsection{\@startsection{subsection}{2}{\z@}{+2.5ex plus +1ex
minus +.2ex}{1.5ex plus .2ex}{\normalsize\it}}
\long\def\@caption#1[#2]#3{\par\addcontentsline{\csname
  ext@#1\endcsname}{#1}{\protect\numberline{\csname
  the#1\endcsname}{\ignorespaces #2}}\begingroup
    \@parboxrestore\small\setlength{\baselineskip}{12pt}
    \@makecaption{\csname fnum@#1\endcsname}{\ignorespaces #3}\par
  \endgroup}
\def\@biblabel#1{#1.~}

\expandafter\ifx\csname mathrm\endcsname\relax\def\mathrm#1{{\rm #1}}\fi
 
\makeatletter

\newcount\@tempcntc
\def\@citex[#1]#2{\if@filesw\immediate\write\@auxout{\string\citation{#2}}\fi
  \@tempcnta\z@\@tempcntb\m@ne\def\@citea{}\@cite{\@for\@citeb:=#2\do
    {\@ifundefined
       {b@\@citeb}{\@citeo\@tempcntb\m@ne\@citea
        \def\@citea{,\penalty\@m\ }{\bf ?}\@warning
       {Citation `\@citeb' on page \thepage \space undefined}}%
    {\setbox\z@\hbox{\global\@tempcntc0\csname
b@\@citeb\endcsname\relax}%
     \ifnum\@tempcntc=\z@ \@citeo\@tempcntb\m@ne
       \@citea\def\@citea{,\penalty\@m}
       \hbox{\csname b@\@citeb\endcsname}%
     \else
      \advance\@tempcntb\@ne
      \ifnum\@tempcntb=\@tempcntc
      \else\advance\@tempcntb\m@ne\@citeo
      \@tempcnta\@tempcntc\@tempcntb\@tempcntc\fi\fi}}\@citeo}{#1}}

\def\@citeo{\ifnum\@tempcnta>\@tempcntb\else\@citea
  \def\@citea{,\penalty\@m}%
  \ifnum\@tempcnta=\@tempcntb\the\@tempcnta\else
   {\advance\@tempcnta\@ne\ifnum\@tempcnta=\@tempcntb \else
\def\@citea{--}\fi
    \advance\@tempcnta\m@ne\the\@tempcnta\@citea\the\@tempcntb}\fi\fi}

\def\beq{\begin{equation}}
\def\eeq{\end{equation}}
\def\beqar{\begin{eqnarray}}
\def\eeqar{\end{eqnarray}}
\def\barr#1{\begin{array}{#1}}
\def\earr{\end{array}}
\def\bfi{\begin{figure}}
\def\efi{\end{figure}}
\def\btab{\begin{table}}
\def\etab{\end{table}}
\def\bce{\begin{center}}
\def\ece{\end{center}}
\def\nn{\nonumber}

\def\text{\textstyle}

\def\al{\alpha}

\def\de{\delta}
\def\eps{\varepsilon}

\def\si{\sigma}

\def\Ga{\Gamma}
\def\De{\Delta}

\def\refeq#1{\mbox{Eq.~\ref{#1}}}

\def\ucite#1{$^{\cite{#1}}$}
\def\citere#1{\mbox{Ref.~\cite{#1}}}

\def\solid{\raise.9mm\hbox{\protect\rule{1.1cm}{.2mm}}}
\def\dash{\raise.9mm\hbox{\protect\rule{2mm}{.2mm}}\hspace*{1mm}}

\newcommand{\GeV}{\unskip\,\mathrm{GeV}}
\newcommand{\MeV}{\unskip\,\mathrm{MeV}}
\newcommand{\TeV}{\unskip\,\mathrm{TeV}}

\def\mathswitchr#1{\relax\ifmmode{\mathrm{#1}}\else$\mathrm{#1}$\fi}

\newcommand{\PW}{\mathswitchr W}
\newcommand{\PZ}{\mathswitchr Z}

\newcommand{\PH}{\mathswitchr H}

\newcommand{\Pb}{\mathswitchr b}
\newcommand{\Pc}{\mathswitchr c}
\newcommand{\Pt}{\mathswitchr t}

\newcommand{\PWpm}{\mathswitchr {W^\pm}}
\newcommand{\PWO}{\mathswitchr {W^0}}

\newcommand{\Rb}{R_\Pb}
\newcommand{\Rc}{R_\Pc}
\newcommand{\Gb}{\Ga_\Pb}
\newcommand{\Gc}{\Ga_\Pc}

\newcommand{\Gh}{\Ga_{\mathrm h}}
\newcommand{\GT}{\Ga_{\mathrm T}}
\newcommand{\Gl}{\Ga_{\mathrm l}}

\def\mathswitch#1{\relax\ifmmode#1\else$#1$\fi}

\newcommand{\MW}{\mathswitch {M_\PW}}
\newcommand{\MWpm}{\mathswitch {M_\PWpm}}
\newcommand{\MWO}{\mathswitch {M_\PWO}}

\newcommand{\MZ}{\mathswitch {M_\PZ}}
\newcommand{\MH}{\mathswitch {M_\PH}}

\newcommand{\Mb}{\mathswitch {m_\Pb}}
\newcommand{\Mt}{\mathswitch {m_\Pt}}

\newcommand{\scrs}{\scriptscriptstyle}
\newcommand{\sw}{\mathswitch {s_{\scrs\PW}}}

\newcommand{\swbar}{\mathswitch {\bar s_{\scrs\PW}}}

\newcommand{\GF}{\mathswitch {G_\mu}}

\newcommand{\chidof}{\chi^2_{\mathrm{min}}/_{\mathrm{d.o.f.}}}

\newcommand{\yb}{y_\Pb}

\newcommand{\alpz}{\alpha(\MZ^2)}
\newcommand{\alps}{\alpha_{\mathrm s}}
\newcommand{\alpsz}{\alpha_{\mathrm s}(\MZ^2)}

\newcommand{\bos}{{\mathrm{bos}}}
\newcommand{\fer}{{\mathrm{ferm}}}

\newcommand{\SC}{{\mathrm{SC}}}
\newcommand{\IB}{{\mathrm{IB}}}

\newcommand{\LEP}{{\mathrm{LEP}}}
\newcommand{\SLD}{{\mathrm{SLD}}}

\def\draftdate{\relax}
\def\mda{\relax}
\def\mua{\relax}
\def\mla{\relax}
\def\draft{
\def\thtystars{******************************}
\def\sixtystars{\thtystars\thtystars}
\typeout{}
\typeout{\sixtystars**}
\typeout{* Draft mode!
         For final version remove \protect\draft\space in source file *}
\typeout{\sixtystars**}
\typeout{}
\def\draftdate{\today}
\def\mua{\marginpar[\boldmath\hfil$\uparrow$]%
                   {\boldmath$\uparrow$\hfil}%
                    \typeout{marginpar: $\uparrow$}\ignorespaces}
\def\mda{\marginpar[\boldmath\hfil$\downarrow$]%
                   {\boldmath$\downarrow$\hfil}%
                    \typeout{marginpar: $\downarrow$}\ignorespaces}
\def\mla{\marginpar[\boldmath\hfil$\rightarrow$]%
                   {\boldmath$\leftarrow $\hfil}%
                    \typeout{marginpar: $\leftrightarrow$}\ignorespaces}
\def\Mua{\marginpar[\boldmath\hfil$\Uparrow$]%
                   {\boldmath$\Uparrow$\hfil}%
                    \typeout{marginpar: $\Uparrow$}\ignorespaces}
\def\Mda{\marginpar[\boldmath\hfil$\Downarrow$]%
                   {\boldmath$\Downarrow$\hfil}%
                    \typeout{marginpar: $\Downarrow$}\ignorespaces}
\def\Mla{\marginpar[\boldmath\hfil$\Rightarrow$]%
                   {\boldmath$\Leftarrow $\hfil}%
                    \typeout{marginpar: $\Leftrightarrow$}\ignorespaces}
\overfullrule 5pt
\oddsidemargin -15mm
\marginparwidth 29mm
}

\makeatletter

\def\eqnarray{\stepcounter{equation}\let\@currentlabel=\theequation
\global\@eqnswtrue
\global\@eqcnt\z@\tabskip\@centering\let\\=\@eqncr
$$\halign to \displaywidth\bgroup\hskip\@centering
  $\displaystyle\tabskip\z@{##}$\@eqnsel&\global\@eqcnt\@ne
  \hskip 2\arraycolsep \hfil${##}$\hfil
  &\global\@eqcnt\tw@ \hskip 2\arraycolsep $\displaystyle\tabskip\z@{##}$\hfil
   \tabskip\@centering&\llap{##}\tabskip\z@\cr}
\def\appendix{\par
 \setcounter{section}{0} \setcounter{subsection}{0}
 \def\thesection{\Alph{section}}}

\makeatother

\newcommand{\lsim}
{\;\raisebox{-.3em}{$\stackrel{\displaystyle <}{\sim}$}\;}
\newcommand{\gsim}
{\;\raisebox{-.3em}{$\stackrel{\displaystyle >}{\sim}$}\;}

\begin{document}

\thispagestyle{empty}
 
\def\thefootnote{\fnsymbol{footnote}}
\setcounter{footnote}{1}
\null
\hfill BI-TP 96/49 \\
\null
\vskip 1.5cm
\begin{center}
{\large ELECTROWEAK THEORY CONFRONTING PRECISION DATA}%
\footnote{Invited Talk presented at the International School of Subnuclear 
Physics, 34th Course: Effective Theories and Fundamental Interactions, 
Erice, Sicily, 3-12 July, 1996.}%
\footnote{Supported by Bundesministerium f\"ur Bildung und Forschung, Germany
and the EC-network contract CHRX-CT94-0579.}
\\
\vskip 3.0em
{\large DIETER SCHILDKNECHT}
\\
\vskip .5em
{\it Fakult\"at f\"ur Physik, Universit\"at Bielefeld \\
Postfach 10 01 31, D-33501 Bielefeld, Germany}
\vskip 2em
\end{center} 
\vskip 1.5cm
\vfil
\centerline{\large ABSTRACT}
\medskip
{\small
We review the empirical evidence for the validity of the Standard Electroweak
Theory in nature. The experimental data are interpreted in terms of an
effective Lagrangian for Z~physics, allowing for potential sources of SU(2)
violation and containing the predictions of the 
Standard Electroweak Theory as a special
case. Particular emphasis is put on discriminating loop corrections due to
fermion-loop vector-boson propagator corrections on the one hand, from
corrections depending on the non-Abelian structure and the Higgs sector on the
other hand. Results from recently obtained fits of the Higgs-boson mass are
reported, yielding $\MH \lsim 550$ GeV [800 GeV] at 95\% C.L.\ 
based on the input
of $\swbar^2(\LEP+\SLD)~ = 0.23165 \pm 0.00024$ [$\swbar^2(\LEP)~=0.23200
\pm 0.00027$]. With respect to LEP2, it is emphasized that first direct
experimental evidence for non-zero non-Abelian couplings among the vector bosons
can be obtained even with restricted $e^+e^-$ luminosity.  
}

\null
\setcounter{page}{0}

\clearpage

\section{Z Physics}

The spirit in which I will look at the electroweak precision data may be
characterized by quoting Feynman who once said:

\begin{quote}
'' In any event, it is always a good idea to try to see how much 
or how little
of our theoretical knowledge actually goes into the 
ana\-lysis of those situations 
which have been experimentally checked.''

\hfill
R.P. Feynman\ucite{FEY} 
\end{quote}

\subsection{The $\alpha(0)$-Born Prediction}
 
The quality of the data on electroweak interactions may be particularly well appreciated by
starting with an analysis in terms of the Born approximation
of the Standard Electroweak Theory (Standard Model, SM)\ucite{GLA,WEI}.
{}From the input of 
\begin{eqnarray}
\alpha (0)^{-1} & = & 137.0359895(61), \nn \\
G_\mu & = & 1.16639(2) \cdot 10^{-5} {\rm GeV}^{-2}, \nn\\
\MZ & = & 91.1863 \pm 0.0020 {\rm GeV}, 
\label{1}
\end{eqnarray}
one may predict the partial width of the Z for decay into leptons, $\Gl$,
the weak mixing angle, $\swbar^2$, and the $W$ mass, $\MW$. 
The Born approximation, more precisely the $\alpha (0)$-Born approximation, in
distinction from the $\alpha (\MZ^2)$-Born approximation to be introduced below,
\begin{eqnarray}
\swbar^2(1-\swbar^2) & = & \frac{\pi \alpha (0)}{\sqrt 2 G_\mu \MZ^2},\nn\\
\Gl & = & \frac{G_\mu M^3_Z}{24\pi \sqrt 2} \left( 1 + (1 - 4
\swbar^2)^2\right),\nn\\
\MW^2 & = & \MZ^2 (1 - \swbar^2) ,  
\label{2}
\end{eqnarray}
then yields  
\begin{eqnarray}
\swbar^2  & = & 0.2121,\nn \\
\Gl &=& 84.85~{\rm MeV}, \nn\\
\MW &=& 80.940~{\rm GeV}    . 
\label{3}
\end{eqnarray}
A comparison with
the experimental data\footnote{Compare the lecture by Martin 
Pohl, these Proceedings.} 
from tab.~1, 
\begin{eqnarray}
\swbar^2(\LEP+\SLD) & = 0.23165 \pm 0.00024,  \nn  \\
\Gl & = 83.91 \pm 0.11\MeV, \nn\\
\MW & = 80.356 \pm 0.125\GeV, 
\label{4}
\end{eqnarray}
shows discrepancies between the $\alpha (0)$-Born approximation and the data by
many standard deviations. 
{\doublerulesep 3pt
\btab
\caption[]{
The 1996 precision data, consisting of the LEP 
data\ucite{LEPEWWG9602}, the SLD value\ucite{SLD} for $\swbar^2$, and the world 
average\ucite{UA2} for $\MW$.
The partial widths $\Gl$, $\Gh$, $\Gb$, and $\Gc$ are obtained from the
observables $R = \Gh/\Gl$, $\si_{\mathrm h} =
(12\pi\Gl\Gh)/(\MZ^2\Ga^2_{\mathrm T})$, 
$\Rb =  \Gb/\Gh$,
$\Rc =  \Gc/\Gh$, and  $\GT$ using the given correlation
matrices. The data in the upper left-hand column will be referred to as
``leptonic sector'' subsequently. Inclusion of the data in the upper
right-hand column will be referred to as ``all data''.
If not stated otherwise,
the theoretical predictions will be based on the input parameters
given in the lower left-hand column of the table, where $\alpz$ is taken
from \citere{JEG}, $\alpsz$ results from the event-shape 
analysis\ucite{BET} at LEP, and $\Mt$ represents the direct
Tevatron measurement\ucite{CDF}.}
\vspace*{.3cm}
\bce
\begin{tabular}{|@{}c@{}||c|c|}
\hline
leptonic sector & \multicolumn{2}{c|}{hadronic sector} \\
\hline
\hline
$\Gl = 83.91 \pm 0.11 \MeV$ & 
$R = 20.778 \pm 0.029$ & $\GT = 2494.6 \pm 2.7 \MeV$ \\
\hline
$\swbar^2|_{\LEP} = 0.23200 \pm 0.00027$ & 
$\si_{\mathrm h} = 41.508 \pm 0.056$ &
$\Gh = 1743.6 \pm 2.5 \MeV$ \\
\hline
$\swbar^2|_{\SLD} = 0.23061 \pm 0.00047$ & 
$\Rb = 0.2179 \pm 0.0012$ &
$\Gb = 379.9 \pm 2.2 \MeV$ \\
\hline
$\swbar^2|_{\mathrm{LEP+SLD}} = 0.23165\pm0.00024$ &
$\Rc = 0.1715 \pm 0.0056$ &
$\Gc = 299.0 \pm 9.8 \MeV$ \\ \hline
$\MW = 80.356 \pm 0.125 \GeV$ & \multicolumn{2}{c|}{} \\
\hline \hline
input parameters & \multicolumn{2}{c|}{correlation matrices} \\
\hline \hline
\begin{array}[b]{c}
\MZ = 91.1863 \pm 0.0020 \GeV \\ \hline
\hspace{5pt} \GF = 1.16639 (2) \cdot 10^{-5} \GeV^{-2} \hspace{5pt} \\ \hline
\alpz^{-1} = 128.89 \pm 0.09 \\ \hline
\alpsz = 0.123\pm0.006 \\ \hline
\Mb = 4.7\GeV \hspace{1pt} \\ \hline
\Mt = 175 \pm 6 \GeV\\
\end{array}
& \multicolumn{2}{c|}{
\begin{array}[b]{|c||c|c|c|c|}
\hline
& \si_{\mathrm h} & R & \GT \\ \hline\hline
\si_{\mathrm h} & \phantom{-}1.00 & \phantom{-}0.15 & -0.14 \\ \hline
R               & \phantom{-}0.15 & \phantom{-}1.00 & -0.01 \\ \hline
\GT             & -0.14 & -0.01 & \phantom{-}1.00 \\ \hline
\earr
\quad
\begin{array}[b]{|c||c|c|c|}
\hline
& \Rb & \Rc \\ \hline\hline
\Rb  & \phantom{-}1.00 & -0.23 \\ \hline
\Rc  & -0.23 & \phantom{-}1.00 \\ \hline
\earr} \\ \hline
\end{tabular}
\ece
\label{tab:data}
\etab
}
  
\subsection{The $\alpha(\MZ^2)$-Born, the Full Fermion-Loop 
and the Complete One-Loop Standard Model Predictions}

Turning to corrections to the $\alpha (0)$-Born approximation, I follow the 1988
strategy
``to isolate and to test
directly the 'new physics' of boson loops and other new phenomena by comparing
with and looking for deviations from the predictions of the
dominant-fermion-loop results''\ucite{GOU}. Accordingly, let us strictly
discriminate\ucite{kn91,BKK,DKK,DKS,DSW} 
vacuum-polarization contributions due to fermion loops in the
photon, $Z$ and $W$ propagators from all other loop corrections, the
``bosonic'' loops, which contain virtual vector bosons within the loops. I note
that this distinction between two classes of loop corrections is gauge invariant
in the $SU(2)_L \times U(1)_Y$ electroweak theory. Otherwise the 
theory would fix the number of
fermion families. The reason for systematically
discriminating fermion loops in the propagators from the rest is in fact
obvious. The fermion-loop effects, leading to ``running'' of coupling constants
and to mixing among the neutral vector bosons, can be precisely predicted from
the {\it empirically known couplings} of the leptons and the (light) quarks,
while other loop effects, such as vacuum polarization due to boson pairs and
vertex corrections, depend on the {\it empirically unknown couplings} among the
vector bosons and the properties of the Higgs scalar. It is in fact the
difference between the fermion-loop predictions and the full one-loop results
which sets the scale\ucite{GOU} 
for the precision required for empirical tests of the
electroweak theory beyond (trivial) fermion-loop effects. One should remind
oneself that the experimentally unknown bosonic interactions are right at the
heart of the celebrated renormalizability properties\ucite{THO} of the electroweak
non-Abelian gauge theory\ucite{WEI}. 

When considering fermion loops, let us first of all look at the contributions of
leptons and quarks to the photon propagator. Vacuum polarization due to leptons
and quarks, or rather hadrons in the latter case, leads to the well-known
increase (``running'') of the electromagnetic coupling as a function of the
scale, at which it is measured. While the contribution of the leptons can be
calculated in a straightforward manner, the contributions due to quarks are more
reliably obtained from the cross section for $e^+ e^-$ annihilation into hadrons
via a dispersion relation\ucite{JEG,JEG1}. As a consequence of the 
experimental errors in the
cross section for $e^+ e^-$ annihilation, in particular in the region below about
3.5 GeV, the value of the electromagnetic fine-structure constant at the $Z$
scale, relevant for LEP1 physics, contains a non-negligible error,   
\beq 
\alpha (\MZ^2)^{-1} = 128.89 \pm 0.09.
\label{5}
\eeq
Replacing $\alpha (0)$ in  
\refeq{2} by $\alpha (\MZ^2)$ implies replacing the
electroweak mixing angle in \refeq{2} by $s^2_0$,
\beq
s^2_0 (1 - s^2_0) = \frac{\pi \alpha (\MZ^2)}{\sqrt 2 G_\mu \MZ^2}, 
\label{6}
\eeq
which may be expected to be a more appropriate parameter  
for electroweak physics at the Z-boson scale
scale than the mixing angle from the $\alpha(0)$-Born approximation of \refeq{2}. 
As the transition from $\alpha (0)$ to $\alpha (\MZ^2)$ is an
effect purely due to the electromagnetic interactions of leptons and quarks (hadrons),
even present in the absence of weak interactions, the relations in \refeq{2} with the
replacement $s^2_W \rightarrow s^2_0$ from \refeq{6} 
may appropriately be called the ``$\alpha
(\MZ^2)$-Born approximation''\ucite{NOV} of the electroweak theory. 

\begin{figure}
\begin{center}
\begin{picture}(15,13)
\put(0.7,6.5){$\bar\sw^2$}
\put(2,0.8){$\MWpm/\MZ$}
\put(11.4,0.6){$\Gl/\MeV$}
\put(-1.8,-6.0){\includegraphics{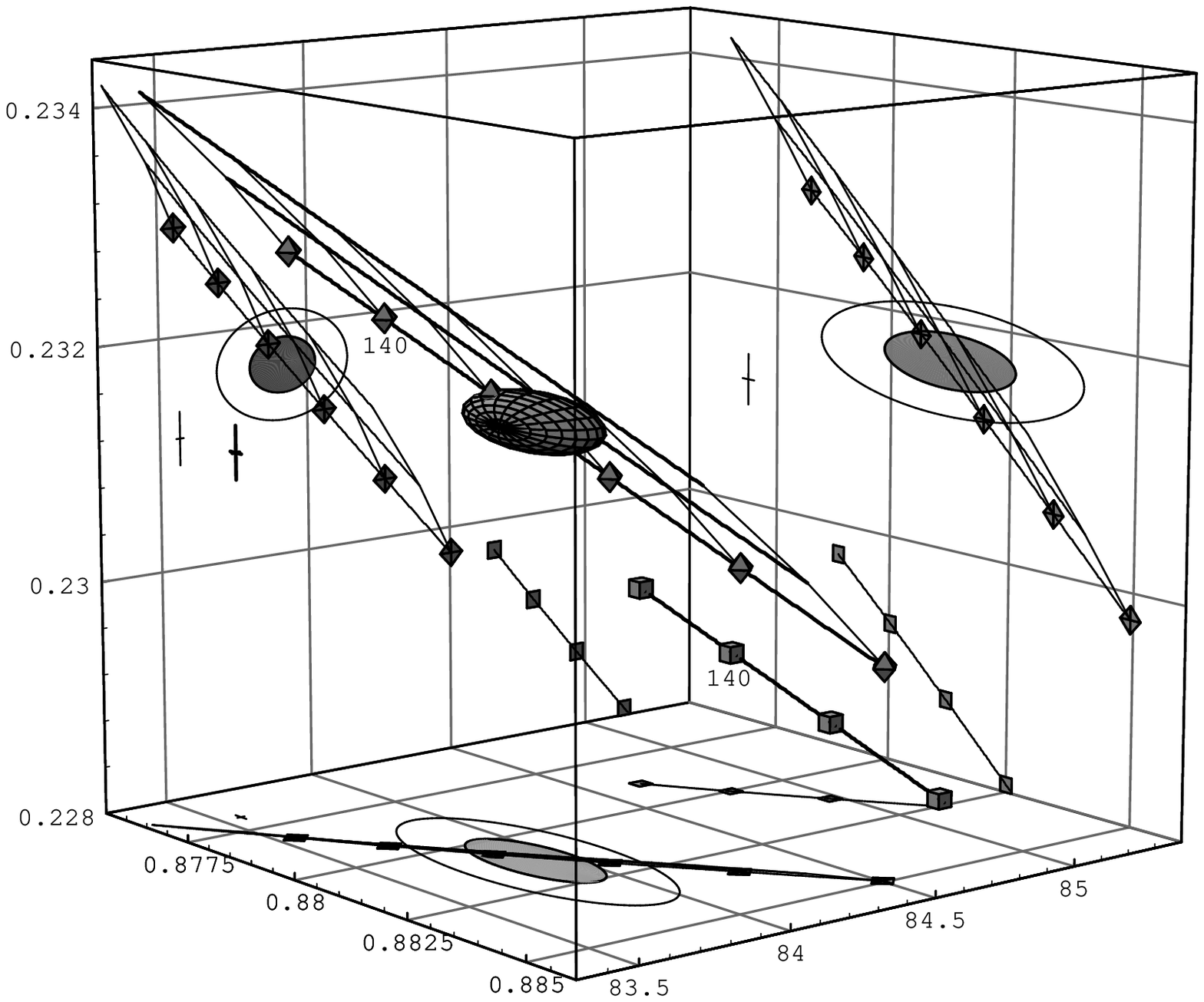}}
\end{picture}
\end{center}
\caption[]{Three-dimensional plot of the $1\sigma$ ellipsoid
of the experimental data in ($\MWpm/\MZ$, $\bar\sw^2$, $\Gl$)-space, using
$\bar\sw^2$ (LEP + SLD) as experimental input for $\swbar^2$, in 
comparison with the full SM prediction (connected lines)
and the pure fermion-loop prediction (single line with cubes).
The full SM prediction is shown for Higgs-boson masses of $\MH = 100
\GeV$ (line with diamonds), $300\GeV$, and 
$1\TeV$ parametrized by $\Mt$
ranging from 120--220$\GeV$ in steps of $20\GeV$. In the pure
fermion-loop prediction the cubes also indicate steps in $\Mt$ of
$20\GeV$ starting with $\Mt = 120$ $\GeV$.
The cross outside the ellipsoid indicates the $\al(\MZ^2)$-Born
approximation with the corresponding error bars,
which also apply to all other theoretical predictions
(1996 update from \citere{DSW}). Note that in the projections on the planes also
the $2\sigma$ contours are shown.  }
\label{sm3d}
\efi
Numerically, one finds 
\beqar
s^2_0 & = & 0.23112 \pm 0.00023 ,            \nn \\
\Gl^{(0)} & = & 83.563 \pm 0.012~{\rm MeV} ,    \nn      \\
M^{(0)}_{\PW} & = & 79.958 \pm 0.011~{\rm GeV} ,   
\label{6a}
\eeqar 
i.e. a large part of the above discrepancy between 
the predictions in \refeq{3} and the data in 
\refeq{4} is due
to the use of the inappropriate 
value of $\alpha (0)$, instead of $\alpha (\MZ^2)$, as
appropriate for Z~physics. Note that the uncertainty in $s^2_0$, as a
consequence of the error in $\alpha (\MZ^2)$, is as large as the error of 
$\swbar^2$ from the measurements at the Z~resonance (compare \refeq{4} or
tab.~1). 

All other fermion-loop effects are due to fermion loops in the W
propagator (relevant simce $G_\mu$ enters the predictions) and in the Z propagator,
and due to the important effect of $\gamma$Z mixing induced by fermions. 
Light fermions as well as the
top quark accordingly yield important contributions to the ``full fermion-loop prediction''
which includes {\it all} fermion-loop propagator corrections. 

In fig.~1, an update of a figure in \citere{DSW}, 
we show the experimental data from the 
``leptonic sector'', $\bar
s^2_W , \Gl , \MW$, in comparison with the $\alpha (\MZ^2)$-Born
approximation, the full fermion-loop prediction, and the complete one-loop
Standard Model results.

We conclude that\ucite{DSW,DKK},
\begin{itemize}
\item
contributions beyond the $\alpha (\MZ^2)$-Born approximation are needed for
agreement with the data, 
\item
contributions beyond the full fermion-loop predictions, based on $\alpha(\MZ^2)$,  
the
fermion-loop contributions to the $W$ and $Z$ propagators 
and to $\gamma Z$ mixing, and the
top quark effects, are necessary, and provided
\item
by additional contributions involving bosonic loops, dependent on the
non-Abelian couplings and the properties of the Higgs boson. 
\end{itemize}

The question immediately arises of what can be said in more detail about the
various contributions due to fermionic and bosonic loops, leading to the final
agreement between theory and experiment.

\subsection{Effective Lagrangian, $\Delta x, \Delta y, 
\varepsilon , \Delta\yb$ Parameters}
 
This question can be answered by an analysis in terms of the parameters 
$\Delta x, \Delta y$ and $\varepsilon$ which within the framework of an effective 
Lagrangian\ucite{BKK,DKK,DKS} 
specify potential sources of $SU(2)$ violation. 
The {\it ``mass parameter''} $\De x$ is related to $SU(2)$ violation by
the masses of the triplet
of charged and 
neutral (unmixed) vector boson via
\beq
\MW^2 \equiv (1+ \Delta x) M^2_{W^0} \equiv xM^2_{W^0},
\label{7}
\eeq
while the {\it ``coupling parameter''}  
$\Delta y$ specifies $SU(2)$ violation among the $W^\pm$ and $W^0$ couplings 
to fermions,
\beq
g^2_{W^\pm} (0) \equiv M^2_{W^\pm} 4 \sqrt 2 G_\mu = (1 + \Delta y) g^2_{W^0}
(\MZ^2) \equiv y g^2_{W^0} (\MZ^2).
\label{8}
\eeq
{}Finally, the {\it ``mixing parameter''} $\varepsilon$ refers to 
the mixing strength in the neutral vector boson sector and quantifies the
deviation 
of $\swbar^2$ from $e^2 (\MZ^2)/g^2_{W^0} (\MZ^2)$,
\beq
\swbar^2 \equiv \frac{e^2 (\MZ^2)}{g^2_{W^0} (\MZ^2)} (1 - \varepsilon ),
\label{9}
\eeq
thus allowing for an unconstrained mixing strength\ucite{kn91,HS} in the neutral
vector-boson sector. 
The effective Lagrangian incorporating the mentioned sources of $SU(2)$
violation for $W$ and $Z$ interactions with leptons is given by\ucite{DKK,BKK}
\beq
{\cal L}_C = -\frac{1}{2} W^{+\mu\nu}W^-_{\mu\nu}
+ \frac{g_\PWpm}{\sqrt 2} \left( j^+_\mu W^{+\mu} + h.c.\right)
+ \MWpm^2 W^+_\mu W^{-\mu}
\label{1a}
\eeq
and 
\beqar
{\cal L}_N & = & - \frac{1}{4} Z_{\mu\nu} Z^{\mu\nu} +
\frac{1}{2} \frac{\MWO^2}{1-\bar\sw^2(1-\varepsilon)}Z_{\mu}Z^{\mu}
- \frac{1}{4} A_{\mu\nu} A^{\mu\nu} \nn\\
&& - e j_{em}^\mu A_\mu +
\frac{g_\PWO}{\sqrt{1-\bar\sw^2 (1-\varepsilon)}}
\left( j^\mu_3 - \bar\sw^2 j^\mu_{em}\right) Z_\mu.
\label{10a}
\eeqar

{}For the observables $\swbar^2 , \MW$ and $\Gl$,
from \refeq{1a} and \refeq{10a} one obtains
\begin{eqnarray}
\swbar^2 (1 - \swbar^2 ) & =& {{\pi \alpha (\MZ^2)}\over{\sqrt 2 G_\mu \MZ^2}}
{y\over x} (1 - \varepsilon) {1\over{\left( 1 + {\swbar^2\over{1 - \swbar^2}}
\varepsilon \right)}}, \nn  \\
{\MW^2\over \MZ^2} & =& (1 - \swbar^2 ) x \left( 1 + {\swbar^2 \over
{1 - \swbar^2}} \varepsilon \right) , \nn\\
\Gl & =& {{G_\mu M^3_Z}\over{24\pi \sqrt 2}} \left( 1 + (1 - 4 \swbar^2 )^2
\right) {x\over y} \left( 1 - {{3\alpha}\over{4\pi}} \right). 
\label{11}
\end{eqnarray}
{}For $x = y = 1$ (i.e., $\Delta x = \Delta y = 0$) and $\varepsilon = 0$ one
recovers the $\alpha (\MZ^2)$-Born approximation,
$\swbar^2 = s^2_0$, discussed previously. 

The extension\ucite{DKS} of the effective Lagrangian \refeq{10a} 
to interactions of neutrinos and
quarks requires the additional coupling parameters $\Delta y_\nu$ for the
neutrino, $\Delta y_b$ for the bottom quark, and $\Delta y_h$ for the remaining
light quarks. In the analysis of the data, for $\Delta y_\nu$ and $\Delta y_h$
which do not involve the non-Abelian structure of the theory, the SM
theoretical results may be inserted without loss of generality
as far as the guiding principle of separating vector-boson--fermion
interactions from interactions containing non-Abelian couplings is
concerned.

We note that the parameters in our effective Lagrangian are related\ucite{DKS} to the 
parameters $\varepsilon_{1,2,3}$ and $\epsilon_b$, 
introduced\ucite{al93} by isolating the
quadratic $\Mt$ dependence, 
\beqar
\parbox{6cm}{$\eps_1=\De x-\De y+0.2\times 10^{-3},$} &&
\eps_2=-\De y+0.1\times 10^{-3}, \nn\\
\parbox{6cm}{$\eps_3=-\eps+0.2\times 10^{-3},$} &&
\eps_\Pb=-\De\yb/2-0.1\times 10^{-3}.
\label{10b}
\eeqar
Essentially the two sets of parameters only differ in $\varepsilon_1$. As
$\varepsilon_1$ contains a linear combination of $\Delta x$ and $\Delta y$, the 
$\MH$-dependent bosonic corrections in $\Delta x$ are confused with the 
$\MH$-insensitive bosonic corrections in $\Delta y$.
The theoretically interesting, but numerically irrelevant additive terms in
\refeq{10b},
considerably smaller than $1 \times 10^{-3}$, originate from a refinement in the
mixing involved in Lagrangian \refeq{10a} and a corresponding refinement in
\refeq{11}. We refer to the original paper\ucite{DKS} for details. 
 
By linearizing relations \refeq{11} with respect to $\Delta x, \Delta y$ and 
$\varepsilon$ and  
inverting them, $\Delta x , \Delta y$ and $\varepsilon$ may 
be deduced from the experimental data on $\swbar^2, \Gl$ and 
$\MW$.
Inclusion of the hadronic $Z$ observables requires that $\Delta x , \Delta y ,
\varepsilon$ and $\varepsilon_b$ are fitted to the experimental data. Actually, one
finds that the
results for $\Delta x , \Delta y , \varepsilon$ are hardly affected by inclusion of
the hadronic observables.  
On the other hand, $\Delta x , \Delta y ,\varepsilon$ and $\Delta y_b$ may be
theoretically determined 
in the standard electroweak theory 
at the one-loop level, strictly discriminating 
between pure fermion-loop predictions and the rest which contains the unknown 
bosonic couplings. The most recent 1996 update\ucite{DS} of such an 
analysis\ucite{DSW,DKK,DKS} is shown in fig.~2.
\begin{figure}
\begin{center}
\begin{picture}(15,14.5)
\put(-2.0,-4.5) {\includegraphics{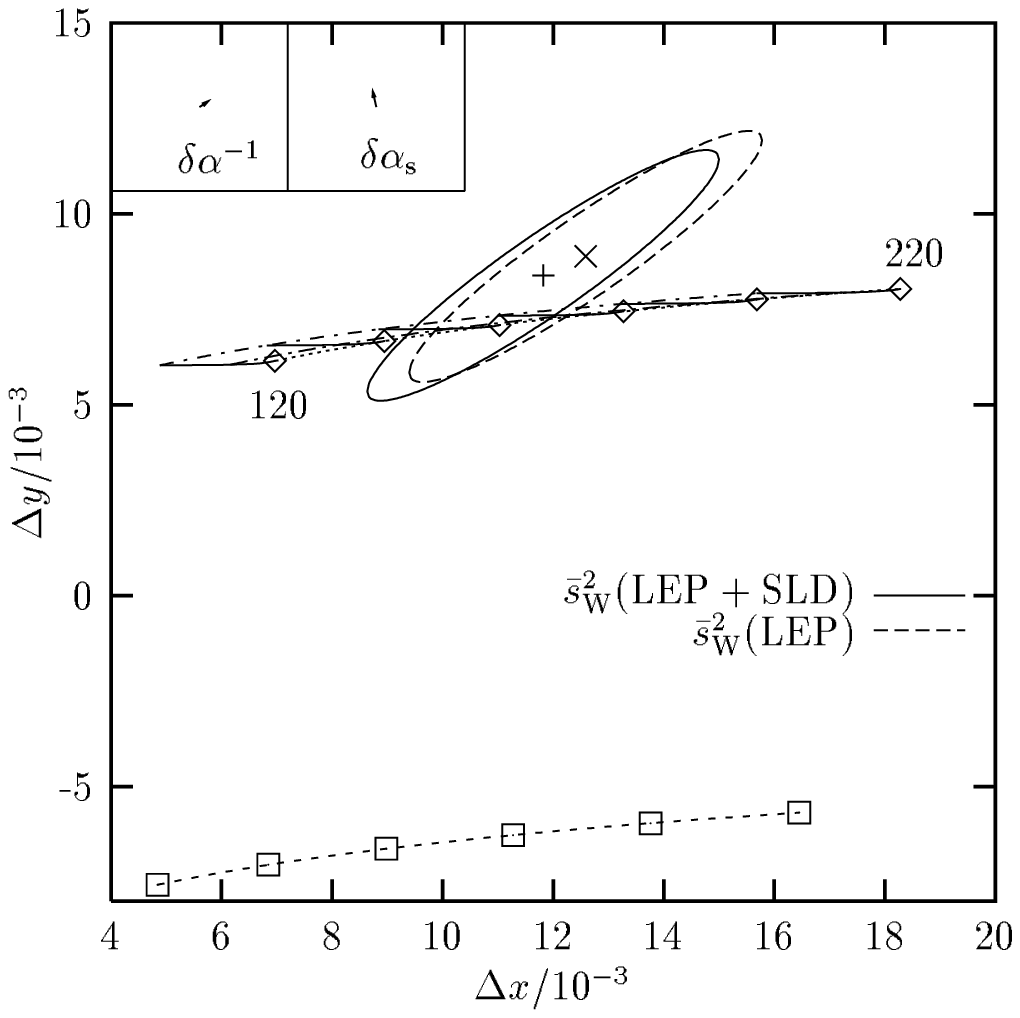}}
\put( 5.7,-4.5) {\includegraphics{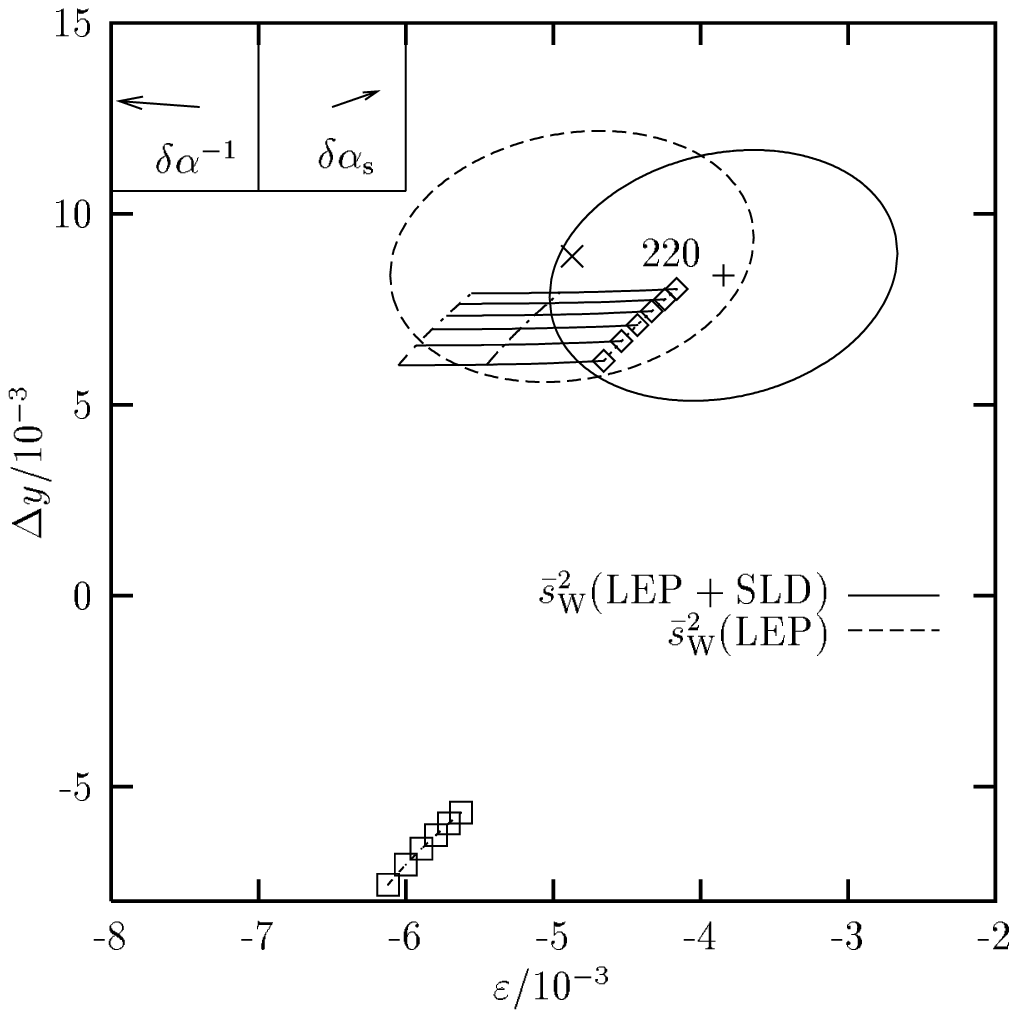}}
\put(-2.0,-12.0){\includegraphics{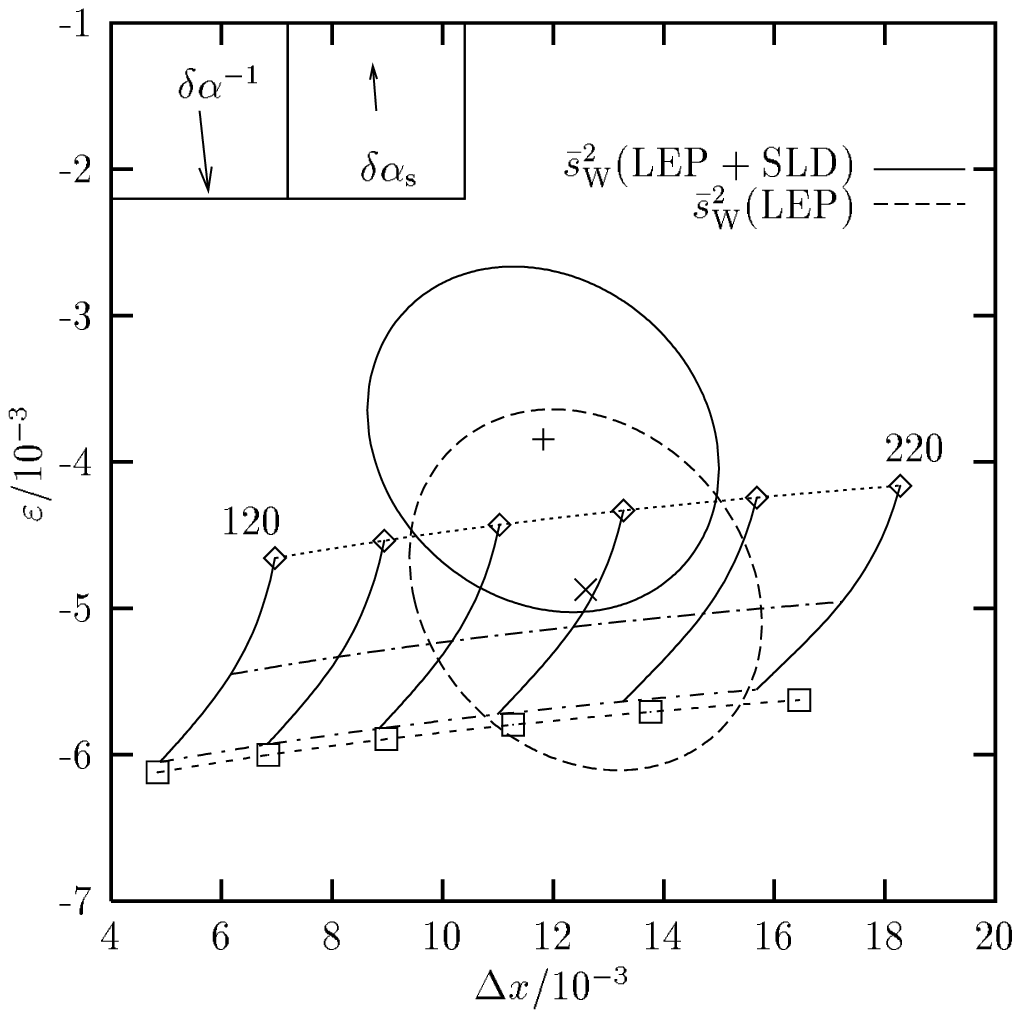}}
\put( 5.7,-12.0){\includegraphics{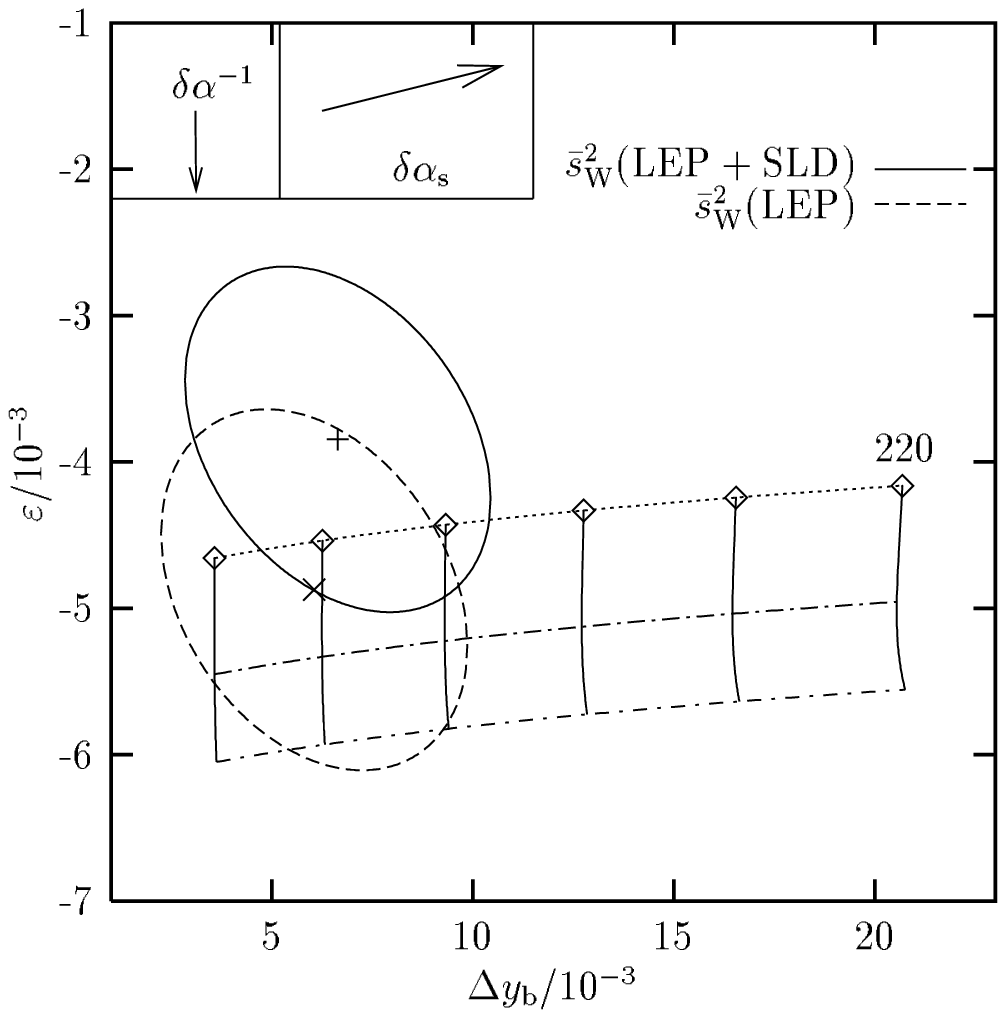}}
\end{picture}
\end{center}
\caption{
The projections of the $1\si$ ellipsoid of 
the electroweak parameters $\De x$, $\De y$, $\eps$,
$\De\yb$ obtained from the 1996 set of data in comparison
with the SM predictions. Both the
results obtained from using $\swbar^2(\LEP)$ and $\swbar^2(\LEP+\SLD)$
as experimental input are shown. The full SM predictions
correspond to Higgs-boson masses of $100\GeV$ (dotted with diamonds), 
$300\GeV$ (long-dashed dotted) and $1\TeV$ (short-dashed dotted)
parametrized by the top-quark mass ranging from
$120\GeV$ to $220\GeV$ in steps of $20\GeV$. 
The pure fermion-loop
prediction is also shown (short-dashed curve with squares) for the
same values of $\Mt$.
The arrows indicate the shifts of the centres of the ellipses upon
changing $\alpz^{-1}$ to $\alpz^{-1}+\de\alpz^{-1}$ and $\alpsz$ to
$\alpsz+\de\alpsz$. (From \citere{DS})}
\label{fig:xyeb}
\efi

According to fig.~2, the data in the $(\varepsilon , \Delta x)$ plane are  
consistent with the theoretical predictions obtained by approximating $\Delta x$
and $\varepsilon$ by their 
pure fermion-loop values, 
\begin{eqnarray}
\Delta x & =& \Delta x_{{\rm ferm}} (\alpha (\MZ^2) , s^2_0 , \Mt^2 \ln \Mt ) 
+ \Delta x_{{\rm bos}} (\alpha (\MZ^2) , s^2_0 , \ln \MH^2) \nn \\
  & & \nn \\
  & \cong & \Delta x_{{\rm ferm}} (\alpha (\MZ^2) , s^2_0 , \Mt^2 ,\ln \Mt) ,  \nn \\
  & &    \nn  \\ 
\varepsilon & = & \varepsilon_{{\rm ferm}} (\alpha (\MZ^2) , s^2_0 , \ln \Mt ) + 
\varepsilon_{{\rm bos}} (\alpha (\MZ^2) , s^2_0 , \ln \MH^2 ) \nn\\
  & & \nn \\
  &\cong & \varepsilon_{{\rm ferm}} (\alpha (\MZ^2) , s^2_0 , \ln \Mt ) . 
\label{12}
\end{eqnarray}
The small contributions of 
$\Delta x_{{\rm bos}}$ and $\varepsilon_{{\rm bos}}$ to $\Delta x$ and $\varepsilon$,
respectively, and the logarithmic dependence on the Higgs mass, $\MH$,
imply the well-known result 
that the data are fairly insensitive to the mass of the Higgs scalar. 
It is instructive to also note the numerical results for $\Delta x_{{\rm ferm}}$
and $\varepsilon_{{\rm ferm}}$, obtained in the Standard Model. They are given by\ucite{DKK}
\beqar
\Delta x_{{\rm ferm}} & = & (2.61 t + 1.34 \log (t) + 0.52) \times 10^{-3} , \nn\\
\varepsilon_{{\rm ferm}} & = & (- 0.45 \log (t) - 6.43) \times 10^{-3} ,
\label{x}
\eeqar
with $t \equiv \Mt^2 / \MZ^2$. The mass parameter $\Delta x$ is dominated by the
$\Mt^2$ term\ucite{VELT} due to weak isospin breaking induced by the top quark, while
$\varepsilon$ is dominated by the constant term due to mixing induced by the light
leptons and quarks. 
 
In distinction from the results for $\Delta x$ and $\varepsilon$, 
where the fermion loops by themselves are consistent with the data, 
a striking effect appears in the plots showing $\Delta y$. 
The theoretical predictions are
clearly inconsistent with the data, unless the fermion-loop contributions to $\Delta y$ 
(denoted by lines with small squares) are supplemented by an 
additional term, which in the standard electroweak theory is due to bosonic effects, 
\beq
\Delta y = \Delta y_{{\rm ferm}} (\alpha (\MZ^2) , s^2_0 , \ln \Mt ) + \Delta y_{{\rm 
bos}} (\alpha (\MZ^2) , s^2_0). 
\label{13}
\eeq
Remembering that $\Delta y$, according to \refeq{8}, 
relates the coupling of the charged boson, $W^\pm$, to leptons as measured in 
$\mu^\pm$ decay, to the coupling of the neutral member, $W^0$, of the vector-boson
triplet at the scale $\MZ \sim \MW$, 
it is not surprising that $\Delta y_{{\rm bos}}$ contains vertex and box corrections
originating from $\mu^\pm$ decay as well as vertex corrections at the $W^0
f\bar f$ ($Z f \bar f$) vertex.
While $\Delta y_{{\rm bos}}$ obviously depends on the trilinear couplings among the vector
bosons, it is insensitive to the Higgs mass, $\MH$. 
 
The experimental data have accordingly become accurate enough to  
isolate loop effects which are insensitive to $\MH$, but depend on the 
self-interactions of the vector bosons, in particular on the trilinear
non-Abelian couplings 
entering the $W f\bar f^\prime$ and $W^0 f\bar f$ ($Z f \bar f)$ 
vertex corrections.

With respect to the interpretation of the coupling parameter, 
$\Delta y$, one further step\ucite{DSW} may appropriately be taken. Introducing the 
coupling of the $W$ boson to leptons, $g_{W^\pm} (\MW^2)$, as defined by the
leptonic $W$-boson width, 
in addition to the previously used low-energy coupling, $g_{W^\pm} (0)$, defined
by the Fermi constant in \refeq{8},
\beq 
\Gamma^W_l = g^2_{W^\pm} (\MW^2) \frac{\MW}{48\pi} \left( 1 + c^2_0
\frac{3\alpha}{4\pi} \right),
\label{14}
\eeq
the coupling parameter, $\Delta y$, in linear 
approximation may be split
into two additive terms, 
\beq
\Delta y = \Delta y^{\SC} + \Delta y^{\IB}.
\label{15}
\eeq
While $\Delta y^{\SC}$ (where ``$\SC$'' stands for ``scale change'') furnishes the
transition from $g_{W^\pm} (0)$ to $g_{W^\pm} (\MW^2)$, 
\beq
g^2_{W^\pm} (0) = (1 + \Delta y^{\SC}) g^2_{W^\pm} (\MW^2),
\label{16}
\eeq
the parameter $\Delta y^{\IB}$ (where ``$\IB$'' stands for ``isospin breaking'')
relates the charged-current and neutral current couplings at the high-mass
scale, $\MW \sim \MZ$, 
\beq
g^2_{W^\pm} (\MW^2) = ( 1 + \Delta y^{\IB}) g^2_{W^0} (\MZ^2),
\label{17}
\eeq
to each other. Note that $\Delta y^{\SC}$ according to \refeq{14} with \refeq{16}
and \refeq{8} can be uniquely extracted from the observables $\MW , \Gamma^W_l$
together with $G_\mu$. 
\begin{figure}
\begin{center}
\begin{picture}(15,10)
\put(-2.0,-13.7){\includegraphics{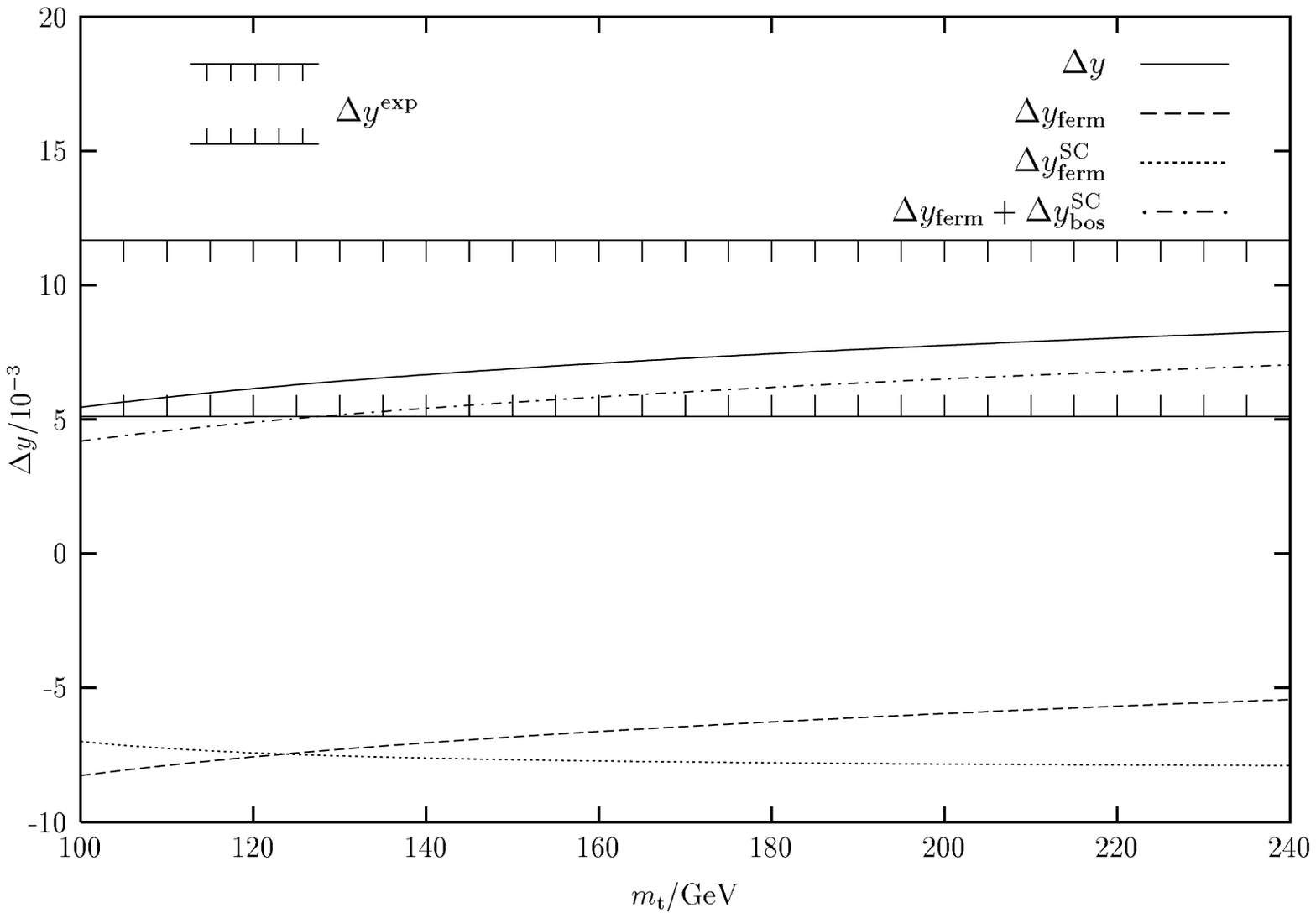}}
\end{picture}
\end{center}
\caption{The one-loop SM predictions for $\De y$, $\De
y_{\fer}$, $\De y^{\SC}_{\fer}$, and $(\De y_{\mathrm
ferm} + \De y^{\SC}_{\bos})$ as a function of $m_\Pt$.
The difference between the curves for $\De y$ and $(\De y_{\fer} +
\De y^{\SC}_{\bos})$ corresponds to the small contribution of $\De
y^{\IB}_{\bos}$.
The experimental value of $\Delta y$,
$\Delta y^{\mathrm exp} = (8.4 \pm 3.3) \times 10^{-3}$, is indicated
by the error band (From \citere{DSW}, 1996 update).}
\label{fig:delysc}
\efi

\btab
\bce
\caption[]{The different contributions (according to \refeq{15})
to the coupling parameter $\Delta y$
(from \citere{DSW}).} 
\vspace*{.3cm}
\begin{tabular}{|l|c|c|c|} \hline
~ & $\Delta y_{{\rm ferm}} \times 10^3$ & $\Delta y_{{\rm bos}} \times 10^3$ & 
$\Delta y \times 10^3$ \\   \hline
SC & $-7.8$ & 12.4 & 4.6  \\ \hline
IB ($\Mt = 175$ GeV) & 1.5 & 1.2 & 2.7  \\   \hline
SC + IB & $-6.3$ & 13.6 & 7.3  \\  \hline  
\end{tabular}
\ece
\etab   

As seen in tab.~2 and fig.~3, the fermion-loop and the bosonic contributions to $\Delta y$ are 
opposite in sign, and both are dominated by their scale-change parts, $\Delta
y^{\SC}$. 
Once, $\Delta y^{\SC}_{{\rm bos}}$ is taken into account, practically no further bosonic 
contributions are needed. 
 
The bosonic loops necessary for agreement
with the data are accordingly recognized as charged-current corrections related
to the use of the low-energy parameter $G_\mu$ in the analysis of the data at
the Z~scale.
Their contribution, due to a gauge-invariant combination of vertex, box and
vacuum-polarization, is opposite in sign and somewhat larger than the
contribution due to fermion-loop vacuum polarization.

Once the input parameters at the Z~scale, $\MZ$ and $\alpha (\MZ^2)$, are
supplemented by the coupling $g_{W^\pm} (\MW^2)$, also defined at the 
scale of $\MW \sim \MZ$ and replacing $G_\mu$, all relevant radiative
corrections are 
contained in $\De x_{\fer}$, $\eps_{\fer}$, and $\De\yb$, and are
either related to weak isospin breaking by the top quark or due
to mixing effects induced by the light leptons and quarks and the top
quark. Compare the numerical results for $\Delta x_{{\rm ferm}}$ and
$\varepsilon_{{\rm ferm}}$ in \refeq{x}.
In addition to $\Delta x_{\rm ferm}$ and $\varepsilon_{\rm ferm}$, 
there is a (small) $\log (\Mt)$ 
isospin-breaking contribution to $\Delta y$ as shown in tab.~2, 
and even an smaller bosonic isospin-breaking contribution. 

In fig.~2, we also show the result for $\Delta y_b$ in the $(\Delta y_b ,
\varepsilon)$ plane. The theoretical prediction for $\Delta y_b$, as a
consequence of a quadratic dependence on $\Mt$, is similar in magnitude to the
one for $\Delta x$. The experimental result for $\Delta y_b$ at the $1\sigma$
level almost includes the theoretical expectation implied by the Tevatron
measurement of $\Mt^{{\rm exp}} = 175 + 6$ GeV. This reflects the fact that the
1996 value of $\Rb$ from tab.~1 is approximately consistent with theory, since
the $\Rb$ enhancement, present in the 1995 data\ucite{LEPEWWG9502} 
has practically gone
away. I will come back to this point when discussing the bounds on $\MH$ implied
by the data. 
 
\subsection{Empirical Evidence for the Higgs Mechanism?}

As the experimental results for $\Delta x$ and $\varepsilon$ are well represented by
neglecting all effects with the exception of fermion loops, and as the bosonic 
contribution to $\Delta y$, which is seen in the data, 
is independent of $\MH$, the 
question as to the role of the Higgs mass and the concept of the Higgs 
mechanism\ucite{HK} with 
respect to precision tests immediately arises.
 
More specifically, one may ask the question whether the experimental results, i.e.
$\Delta x, \Delta y , \varepsilon$, and $\De\yb$ 
can be predicted even without the very concept of the 
Higgs mechanism. 
 
In \citere{DGS} we start from the well-known fact that the standard electroweak 
theory without Higgs particle may credibly be reconstructed\ucite{HS}
within the framework of a 
massive vector-boson theory (MVB) 
with the most general mass-mixing term which preserves 
electromagnetic gauge invariance. This theory is then cast into a form which is 
invariant under local $SU(2) \times U(1)$ transformations by introducing three auxiliary
scalar fields \'a la Stueckelberg\ucite{ST,KG}. 
As a consequence, loop calculations may be carried out
in an arbitrary $R_\xi$ gauge
in close analogy to the SM, even though the non-linear realization of
the $SU(2) \times U(1)$, obviously, does not imply renormalizability of
the theory.
 
Explicit loop calculations show that indeed the Higgs-less observable $\Delta y$, 
evaluated in the MVB, coincides with $\Delta y$ evaluated in 
the standard electroweak theory, i.e.\ in particular for the bosonic part,
we have\footnote{Actually, in the SM there is an additional contribution
of ${\cal O}(1/\MH^2)$ which is irrelevant numerically for 
$\MH \gsim 100$ GeV.
Note that the $\MH$-dependent contributions to interactions violating
custodial SU(2) symmetry turn out to be suppressed\ucite{he94sdcgk} by a
power of $1/\MH^2$ in the SM relative to the expectation from
dimensional analysis.
The absence of a $\log\MH$ term in $\De y$ and the absence of
a $\MH^2\log\MH$ term in $\De x$ in the SM thus appear on equal footing 
from the point of view of custodial SU(2) symmetry. 
In contrast,
no suppression relative to dimensional analysis is present in the 
mixing parameter $\varepsilon$, which does not violate custodial SU(2) 
symmetry.
}
\beq
\Delta y_{{\rm bos}}^{{\rm MVB}} \equiv \Delta y_{{\rm bos}}^{{\rm SM}}.
\label{18}
\eeq
In the case of  $\Delta x_{{\rm bos}}$ and $\varepsilon_{{\rm bos}}$, one finds that the 
MVB and the SM differ by the replacement
$\ln \MH \Leftrightarrow \ln \Lambda $ ,
where $\Lambda$ denotes an ultraviolet cut-off.
{}For $\Lambda \lsim 1$ TeV, accordingly, 
\begin{eqnarray}
\Delta x^{{\rm MVB}} & \cong \Delta x_{{\rm ferm}}^{{\rm MVB}} = \Delta x_{{\rm ferm}}^{{\rm 
SM}}, \nn\\
\varepsilon^{{\rm MVB}} &\cong \varepsilon_{{\rm ferm}}^{{\rm MVB}} = 
\varepsilon_{{\rm ferm}}^
{{\rm SM}}.
\label{19}
\end{eqnarray}
 
In conclusion, the MVB can indeed be evaluated at one-loop level
at the expense of introducing a logarithmic cut-off, $\Lambda$. This cut-off only 
affects $\Delta x$ and $\varepsilon$, whose bosonic contributions cannot be well resolved 
experimentally anyway. 
 
The quantity $\Delta y$, whose bosonic contributions are essential for agreement with 
experiment, is independent of the Higgs mechanism, i.e. it is convergent for
$\Lambda \rightarrow \infty$ in the MVB theory.  
It depends on the non-Abelian couplings
of the vector bosons among each other, which enter the vertex corrections at the 
$W$ and $Z$ vertices. Even though the data cannot discriminate between the 
MVB and the standard model with Higgs scalar, the Higgs mechanism 
nevertheless yields 
the only known simple physical realization of the cut-off $\Lambda$
(by $\MH$) which guarantees renormalizability. 
 
\subsection{Bounds on the Higgs-Boson Mass}

We return to the description of the data in the SM, and
in particular discuss the question, in how far the mass of the Higgs boson,
$\MH$, can be deduced from the precision data. 

In Section 1.3 we noted that the full (logarithmic) dependence on $\MH$ is
contained in the mass parameter, $\Delta x$, and in the mixing parameter,
$\varepsilon$. The experimental restrictions on $\MH$ may accordingly be visualized
by showing the contour of the data in the $(\Delta x, \varepsilon )$ plane for the
fixed (theoretical) value of $\Delta y \cong 7\times 10^{-3}$ (corresponding to $\Mt =
175 \pm 6$ GeV) in comparison with the $\MH$-dependent theoretical predictions for
$\Delta x$ and $\varepsilon$. Fig.~4 illustrates the delicate dependence of 
bounds
for $\MH$ on the experimental input for $\swbar^2, \alpha (\MZ^2)$ and
$\Mt^{{\rm exp}}$. The bounds on $\MH$, one can read off from fig.~4, are
qualitatively in agreement of the results of fits to be discussed next.  
\begin{figure}
\begin{center}
\begin{picture}(15,8.2)
\put( 3.0,7.8){$\swbar^2(\LEP+\SLD)$}
\put(10.5,7.8){$\swbar^2(\LEP)$}
\put(-2.0,-12.7){\includegraphics{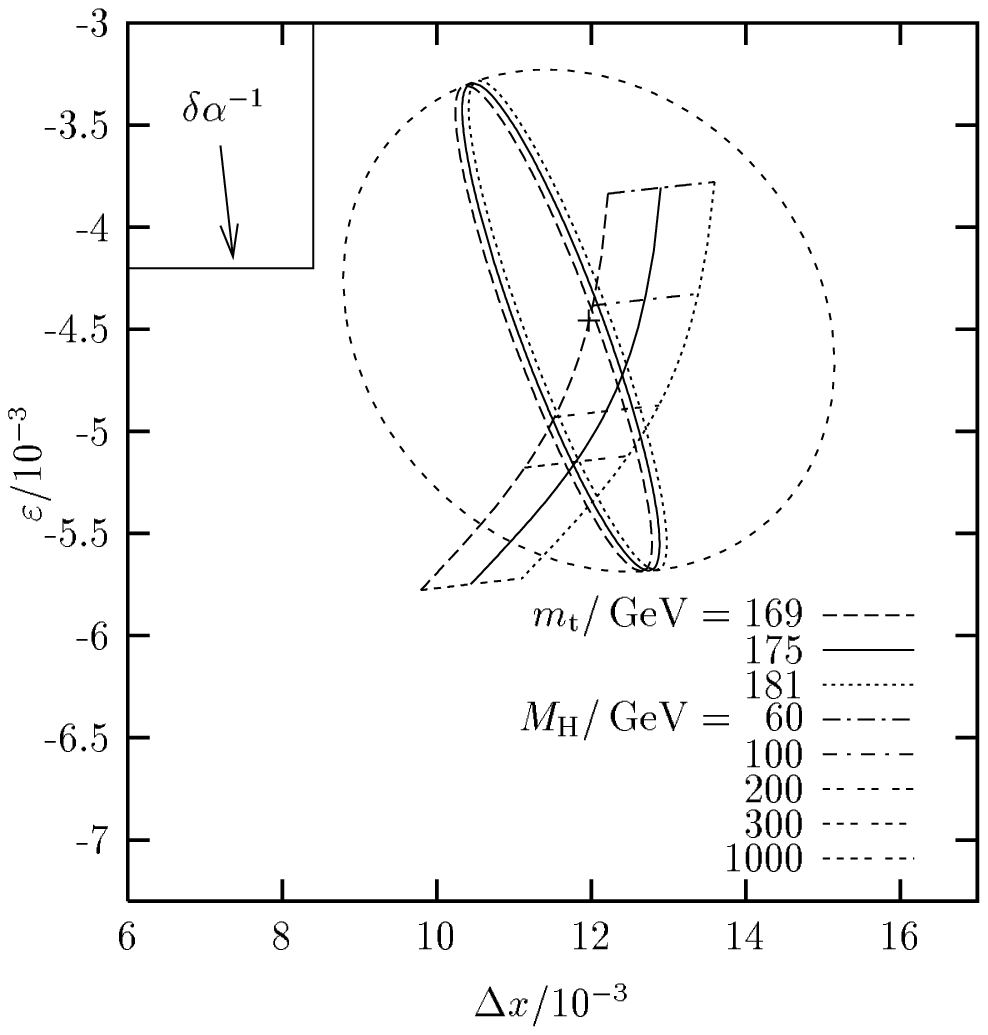}}
\put( 4.7,-12.7){\includegraphics{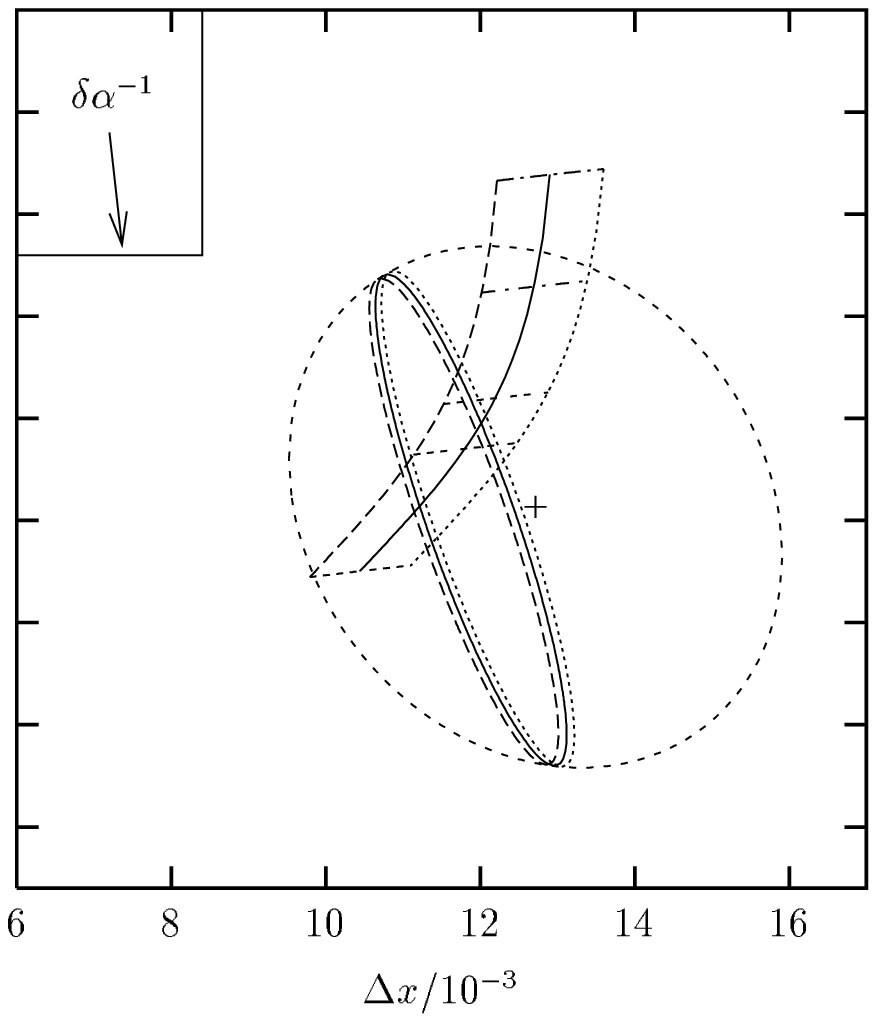}}
\end{picture}
\end{center}
\caption{The
$1\sigma$ contour of the experimental data in the
$(\Delta x,\varepsilon)$
plane defined by $\Delta y \protect\cong 7 \times 10^{-3}$
(corresponding to $\Mt = 175 \pm 6$ GeV). The cut of the contour with the
theoretical predictions for $\Mt = 175 \pm 6$ GeV yields the experimental
bounds on $\MH$. The projection of the data ellipsoid on the 
$(\Delta x,\varepsilon)$ plane, also shown, differs slightly from the one 
in fig.~2,
since the data from the leptonic sector only were used for the present figure.
}
\efi

Precise bounds on $\MH$ require a fit to the experimental data. In order to
account for the experimental uncertainties in the input parameters of $\alpha
(\MZ^2), \alps (\MZ^2)$ and $\Mt$, 
four-parameter $(\Mt,
\MH, \alpha (\MZ^2), \alps (\MZ^2))$ fits to various sets of observables from
tab.~1 were actually performed in \citere{DSCW,DS}. 
The Higgs-boson mass, $\MH$, and $\alps (\MZ^2)$ were treated as free fit
parameters, while for $\alpha (\MZ^2)$ and $\Mt$ 
the experimental constraints from tab.~1 were used.

The results of the 1996 update (taken from \citere{DS}) of the 
fits\ucite{DSCW}\footnote{Compare also \citere{LEPEWWG9602,ho96} for 
$\MH$-fits to the 1996
electroweak data, and \citere{mo94,el95} for $\MH$ fits to previous sets 
of data.}
are presented in the plots of
$\Delta \chi^2 \equiv \chi^2 - \chi^2_{\min}$ against $\MH$ of fig.~5.
\begin{figure}
\begin{center}
\begin{picture}(15,13.2)
\put(-2.5,-13.2){\includegraphics{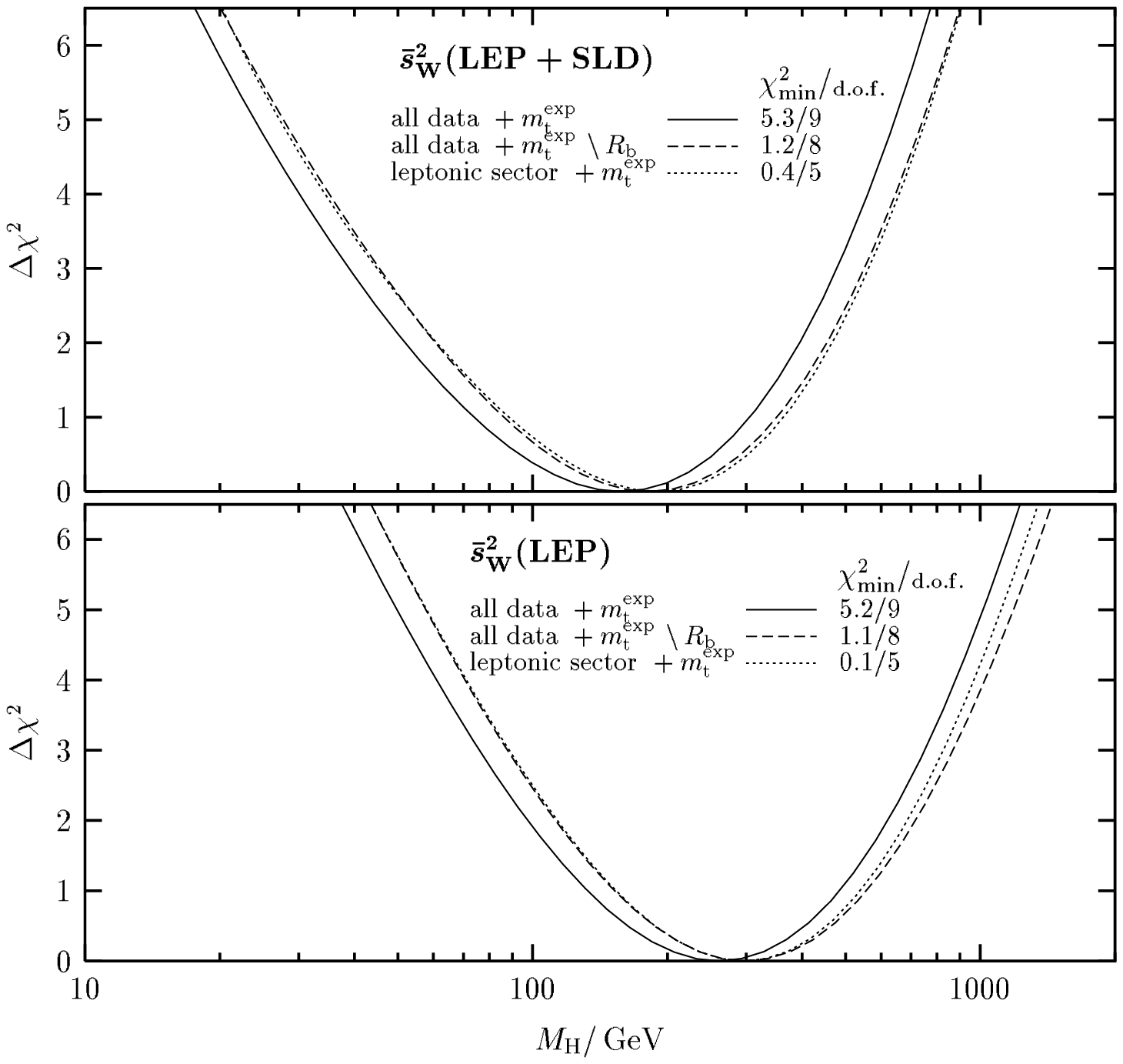}}
\end{picture}
\end{center}
\caption{$\De\chi^2=\chi^2-\chi^2_{\mathrm{min}}$ is plotted against
$\MH$ for the $(\Mt,\MH,\alpz,\alpsz)$ fit to various sets of 
physical observables.
{}For a chosen input for
$\swbar^2$, as indicated, we show the result of a fit to
\protect\\
(i) the full set of 1996 data, 
$\swbar^2$, $\MW$, $\GT$, $\si_{\mathrm{h}}$, $R$, $\Rb$, $\Rc$, 
together with
$\Mt^{\exp}$, $\alpz$,
\protect\\
(ii) the 1996 set of (i) upon exclusion of $\Rb$,
\protect\\
(iii) the 1996 ``leptonic sector'' of $\swbar^2$, $\MW$, $\Gl$, 
together with $\Mt^{\exp}$, $\alpz$. (From \citere{DS})
}
\label{fig:Dchi}
\efi
\begin{figure}
\begin{center}
\begin{picture}(15.1,19.5)
\put(-0.2,-0.7){\includegraphics{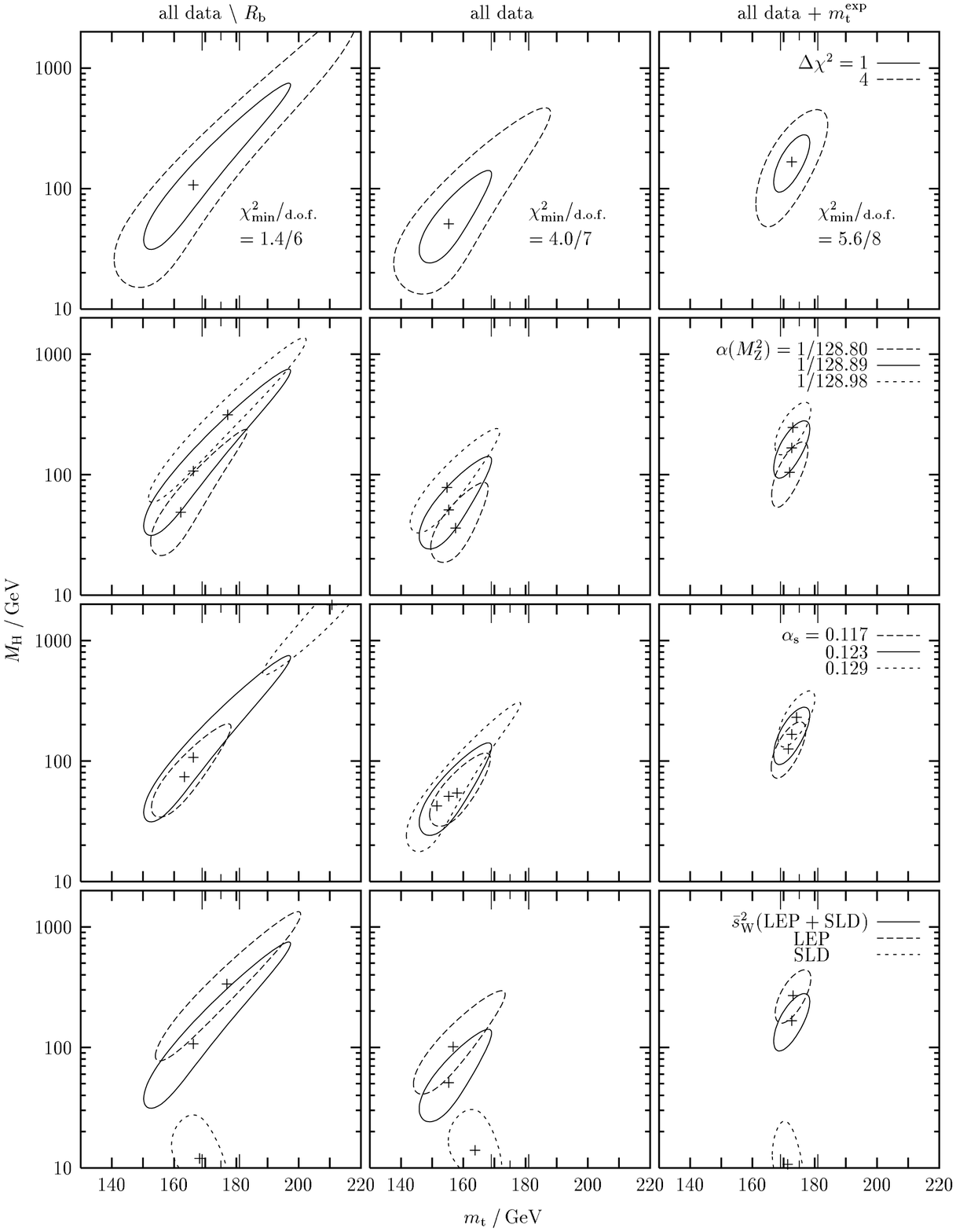}}
\end{picture}
\end{center}
\caption{}
\efi
\addtocounter{figure}{-1}
\begin{figure}
\caption[xxx]{
The results of the two-parameter $(\Mt,\MH)$ fits within the SM
are displayed in the $(\Mt, \MH)$ plane. The different columns refer 
to the sets of experimental data used in the corresponding 
fits, \protect\\
(i) ``all data $\backslash\Rb$'': $\swbar^2(\LEP+\SLD)$, 
$\MW$, $\GT$, $\si_{\mathrm{h}}$, $R$, $\Rc$, 
\protect\\
(ii) ``all data'': $\Rb$ is added to set (i), 
\protect\\
(iii) ``all data + $\Mt^{\exp}$'': $\Rb,\Mt^{\exp}$ are added to the set (i).
\protect\\
The second and third row shows the shift resulting from changing 
$\alpz^{-1}$ and $\alpsz$, respectively,
by one standard deviation in the SM prediction.
The fourth row shows the effect of replacing $\swbar^2(\LEP+\SLD)$ by 
$\swbar^2(\LEP)$ and $\swbar^2(\SLD)$ in the fits.
Note that the $1\sigma$ boundaries given in the first row 
are repeated identically in each row, in order to facilitate comparison with 
other boundaries. 
The value of $\chidof$ given in the plots refers to the 
central values of $\alpz^{-1}$ and $\alpsz$.
In all plots the empirical value 
of $\Mt^{\exp} = 175 \pm 6 \GeV$ is also indicated. (From \citere{DS})}
\label{fig:mtmhfit}
\efi

As $\chi^2_{\min}$ is smallest for the fit to the ``leptonic sector'' of $\bar
s^2_W , \MW , \Gl$ together with $\Mt^{\exp}$, and $\alpha (\MZ^2)$, while the
$1\sigma$ errors are approximately the same in the three fits shown in fig.~5,
we quote the result from the leptonic sector as the most reliable one,
\beqar
\MH = 190^{+174}_{-102} {\rm GeV}, & \quad {\rm using} \quad 
& \swbar^2 ({\rm LEP +
  SLD})_{'96} = 0.23165 \pm 0.00024, \nn  \\
\MH = 296^{+243}_{-143} {\rm GeV}, & \quad {\rm using} \quad 
& \swbar^2 ({\rm LEP})_{'96}
= 0.23200 \pm 0.00027  
\label{20}
\eeqar
based on the 1996 set of data. It implies the $1\sigma$ bounds of 
$\MH \lsim 360$ GeV and $\MH \lsim 540$ GeV,
using $\swbar^2$ (LEP + SLD) and $\swbar^2$ (LEP), respectively, and 
\beqar
\MH \lsim 550 {\rm GeV}~(95\% {\rm C.L.}) & \quad {\rm using} \quad 
& \swbar^2 ({\rm SEP +
  SLD})_{'96}, \nn \\
\MH \lsim 800 {\rm GeV}~(95\% {\rm C.L.}) & \quad {\rm using} \quad 
& \swbar^2 ({\rm LEP})_{'96}.
\label{21}
\eeqar
The fact that the results \refeq{20} and \refeq{21} do not require $\alps
(\MZ^2)$ as input parameter (apart from two-loop effects), and accordingly are
independent of the uncertainties in $\alps (\MZ^2)$, provides an additional reason
for the restriction to the leptonic sector when deriving bounds for
$\MH$. Moreover, we note that according to fig.~5 the results for $\MH$ given by
\refeq{20} and \refeq{21} practically
do not change if the $\alps (\MZ^2)$-dependent
observables, $\GT$ and $\Gh$, the total and hadronic $Z$ widths, are
included in the fit. Inclusion of $\GT$ and $\Gh$ 
provides important information on
$\alps (\MZ^2)$, however. One obtains\ucite{DS} 
$\alps (\MZ^2) = 0.121 \pm 0.003$
and $\alps (\MZ^2) = 0.123 \pm 0,003$ depending on whether $\swbar^2$ (LEP
+ SLD) or $\swbar^2$ (LEP) was used in the fit. Both values are consistent
with the event-shape result given in tab.~1. The impact of also including $\Rb$
in the fit, also shown in fig.~5, will be commented upon below. Inclusion or
exclusion of $\Rc$ is unimportant, as the error in $\Rc$ is considerable. 

As mentioned, the above results on $\MH$ are based on the 1996 set of data which
was presented at the Warsaw International Conference on High Energy Physics 
which took place 
towards the end of July, two weeks after the International School of Subnuclear
Physics in Erice. Two results presented in Warsaw are of particular importance
with respect to the bounds on $\MH$. 

{}First of all, the value of $\Mt = 175 \pm 6$ GeV reported in Warsaw and given in
tab.~1 
is significantly more precise than the 1995 result\ucite{LEPEWWG9502} of $\Mt =
180 \pm 12$ GeV. The decrease in the error on $\Mt$, due to the $(\Mt, \MH)$
correlation in the theoretical predictions for the observables, 
clearly visible in fig.~4, led to a
substantially narrower $\Delta\chi^2$ distribution in fig.~5 compared with the
results based on the 1995 set of data. Indeed, the 1995 leptonic set of
data had implied\ucite{DSCW}
\beqar
\MH & = & 152^{+282}_{-106}\GeV \quad {\rm using} \quad \swbar^2~({\rm
LEP + SLD})_{'95} = 0.23143\pm 0.00028, \nn \\
\MH & = & 353^{+540}_{-224}\GeV \quad {\rm using} \quad \swbar^2~({\rm LEP})_{'95} =
0.23186 \pm 0.0034 , 
\label{22}
\eeqar
i.e., central values similar to the ones in \refeq{20}, but with substantially larger
errors. 

The second and most pronounced change occurred in the result for $\Rb \equiv
\Gamma_b / \Gh$. The enhancement in the 1995 value\ucite{LEPEWWG9502} of 
$\Rb = 0.2219 \pm
0.0017$ of almost four standard deviations with respect to the theoretical
prediction, according to the 1996 result of $\Rb = 0.2179 \pm 0.0012$ 
presented in Warsaw, has reduced to less
than two standard deviations. 
In order to discuss the impact of $\Rb$ on the results for $\MH$, 
if $\Rb$ is included in the fits, we
recall that the theoretical prediction for $\Rb$ is (practically) independent of
the Higgs mass, but significantly dependent on $\Mt$. As the theoretical
prediction for $\Rb$ increases with decreasing mass of the top quark, $\Mt$, an
experimental enhancement of $\Rb$ effectively amounts\ucite{DSCW}
to imposing a low 
top-quark mass in fits of $\Mt$ and $\MH$, as soon as $\Rb$ is included in the
fits. Lowering the top-quark mass in turn implies a lowering of $\MH$ as a
consequence of the $(\Mt, \MH)$ correlation present in the theoretical values of
the other observables. Looking at fig. 5, we see that this effect of lowering
$\MH$ is not very significant with the 1996 value of $\Rb$ and the 1996
error in $\Mt$. The ``$\Rb$-crisis'' in the 1995 data, in contrast, led to a 
substantial decrease in the deduced value of $\MH$ to e.g. $\MH =
81^{+144}_{-52}$ GeV with $\swbar^2$ (LEP + SLD). As stressed in
\citere{DSCW},
this low value of $\MH$ had to be rejected, however, as the effective top-quark
mass induced by including $\Rb$ was substantially below the result from the
direct measurements at the Tevatron. Other consequences from the
``$\Rb$-crisis'', such as an exceedingly low value of $\alps \cong 0.100$ 
required upon allowing for a necessary
non-standard $Zb \bar b$ vertex, as discussed during my lecture in Erice, have also
gone away, and a very satisfactory and consistent overall picture of
agreement with Standard Model predictions has emerged. Speculations on the
existence of a ``leptophobic''\ucite{RE} or a ``hadrophilic'' extra
boson\ucite{RE,ALT,FR}, offered as
potential solutions\footnote{Compare the Erice lectures by Paul Frampton and
Dimitri Nanopoulos, these Proceedings.} 
to the ``$\Rb$-crisis'', do not seem to be realized in
nature. 

The delicate interplay of the experimental results for $\swbar^2 , \Rb$ and
$\Mt$ in constraining $\MH$ and the dependence of $\MH$ on $\alpha (\MZ^2)$ and
$\alps (\MZ^2)$
is visualized in the two-parameter $(\Mt , \MH)$ fits shown in
fig.~6. With its caption, fig.~6 is fairly self-explanatory. For a detailed
discussion we refer to the original papers\ucite{DSCW,DS}.
We only note the considerable dependence of the bounds resulting for $\MH$ on
whether the experimental value for $\Mt$ is included in the fit and the strong
dependence of $\MH$ on a $1\sigma$ variation of $\alpha (\MZ^2)$ and $\alps$.
{}Fig.~6 also shows that the SLD value of $\swbar^2$, when
taken by itself, would rule out an interpretation of the data in terms of the
standard Higgs mechanism, since the resulting Higgs mass, $\MH$, is much
below the lower bound of $\MH \geq 65$ GeV following from the direct Higgs-boson
search at LEP.    

\section{Production of \boldmath{W$^+$W$^-$} at LEP2}

At this meeting we learnt that LEP2 has successfully started running at an
energy above the $\PW^+\PW^-$ production threshold, and the first $W^+W^-$
events were presented. It is appropriate to add a remark on what we can learn
on the couplings of the vector bosons among each other, even with the restricted
luminosity to be accumulated in a few weeks or in a few months of running of LEP~2. 

I start by quoting my dinstinguished  
late friend J.J. Sakurai, who was a frequent lecturer
at the International School of Subnuclear Physics here in Erice. In his
characteristic way of looking at physics, he said\ucite{SAK}:

\begin{quote}
``To quote Weinberg [Rev. Mod. Phys. {\bf 46} (1974) 255]
\begin{quote}
`Indeed, the best way to convince oneself that gauge theories may have something
to do with nature is to carry out some specific calulation and watch the
cancellations before one's very eyes'.
\end{quote} 
Does all this sound convincing? In any
case it would be fantastic to see how the predicted cancellations take place
{\it experimentally} at colliding beam facilities - LEPII? - in the 200 to 300
GeV range.'' 
\end{quote}
Unfortunately, J.J.\ was overly optimistic concerning the energy range of LEP2. 

In connection with the discussion of the coupling parameter $\Delta y$ in
sect.~3, we stressed that the agreement with the LEP1 data at the Z provides
convincing {\it indirect} experimental evidence for the non-Abelian couplings of the
Standard Model. More {\it direct, quantitative} information can be deduced from future
data on $e^+ e^- \rightarrow W^+ W^-$.

My remark will be brief, and essentially consists of showing two figures on the
accuracy which we may expect, when extracting trilinear vector-boson couplings
from measurements of the reaction $e^+ e^- \rightarrow W^+ W^-$ at LEP2. Restricting
ourselves to dimension-four, P- and C-conserving interactions, the general
phenomenological Lagrangian for trilinear vector boson couplings\ucite{BKRS}
\begin{eqnarray}
{\cal L}_{int}&=&-ie[A_\mu(W^{-\mu\nu}W^+_\nu-W^{+\mu\nu}W^-_\nu)
+F_{\mu\nu}W^{+\mu}W^{-\nu}]\nonumber\\
&&-iex_\gamma F_{\mu\nu}W^{+\mu}W^{-\nu}\nonumber\\
&&-ie(\frac{c_W}{s_W}+\delta_Z)[Z_\mu(W^{-\mu\nu}W^+_\nu-W^{+\mu\nu}W^-_\nu)
+Z_{\mu\nu}W^{+\mu}W^{-\nu}]\nonumber\\
&&-iex_Z Z_{\mu\nu}W^{+\mu}W^{-\nu}
\label{23}
\end{eqnarray}
is obtained by supplementing the trilinear interactions of the SM with an
additional anomalous magnetic-moment coupling of strength $x_\gamma$, by
allowing for arbitrary normalization of the Z~coupling via $\delta_Z$, and
by adding an additional anomalous weak magnetic dipole coupling of the Z of
strength $x_Z$.
Compare \citere{gks2} for a representation of the effective Lagrangian \refeq{23}
in an $SU(2) \times U(1)$ gauge-invariant form.  
The SM corresponds to $x_\gamma = \delta_Z = x_Z = 0$.

Non-vanishing values of $x_\gamma$ parametrize deviations of the magnetic
dipole moment, $\kappa_\gamma$, from its SM value of $\kappa_\gamma = 1$, as
according to \refeq{23}, 
\beq 
x_\gamma \equiv \kappa_\gamma - 1.
\label{24}
\eeq
We note that $\kappa_\gamma = 1$ corresponds to a gyromagnetic ratio , $g$,
of the $W$ of magnitude $g = 2$ in units of the particle's Bohr-magneton
$e/2 \MW$, while $\kappa_\gamma = 0$ corresponds to $g = 1$ as obtained for a
classical rotating charge distribution. The weak dipole coupling, $x_Z$, may be
related to $x_\gamma$ by imposing ``custodial'' $SU(2)$ symmetry via\ucite{MSS}
\beq
x_Z = - \frac{s_W}{c_W} x_\gamma, 
\label{25}
\eeq
thus reducing the number of free parameters to two independent ones in \refeq{23}.
Relation \refeq{25} follows from requiring the absence of an $SU(2)$-violating
interaction term solely among the members of the $SU(2)$ triplet, $W^3_{\mu\nu}
W^{+\mu} W^{-\nu}$, when rewriting the Lagrangian in the $BW^3$ base (or the 
$\gamma W^3$ base). This requirement is motivated by the validity of $SU(2)$
symmetry for the vector-boson mass term, i.e. 
from the observation that the deviation of the experimental
value for $\Delta x$ from $\Delta x = 0$ in sect.~1.3 is fully explainable by
radiative corrections, thus ruling out a violation of ``custodial'' $SU(2)$
symmetry by the vector boson masses at a high level of accuracy. 

We also note the relation of $\delta_Z$ to the gauge coupling $\hat g$
describing the trilinear coupling between $W^0$ and $W^\pm$ in the $BW^0$ (or
$\gamma W^3$) base, 
\beq
e \delta_Z \equiv g_{ZWW} - e \frac{c_W}{s_W} = \frac{\hat g}{c_W} -
\frac{e}{s_W c_W} .
\label{30}
\eeq
The SM corresponds to $\hat g = e / s_W$. 

\begin{figure}
\begin{center}
\begin{tabular}{c@{~}c}
\epsfig{file=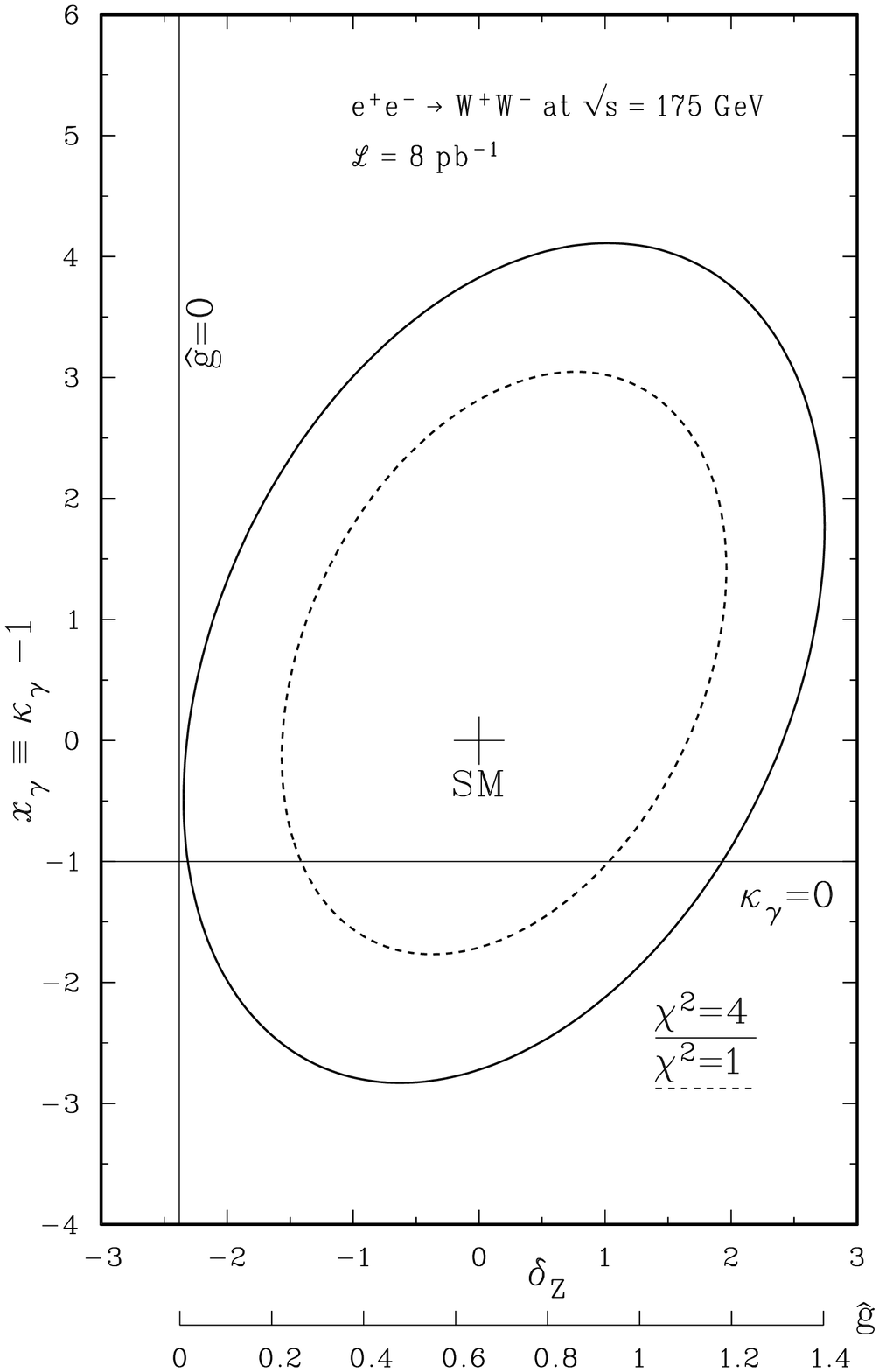,width=7.2cm,
bbllx=40,bblly=40,bburx=490,bbury=710}&
\epsfig{file=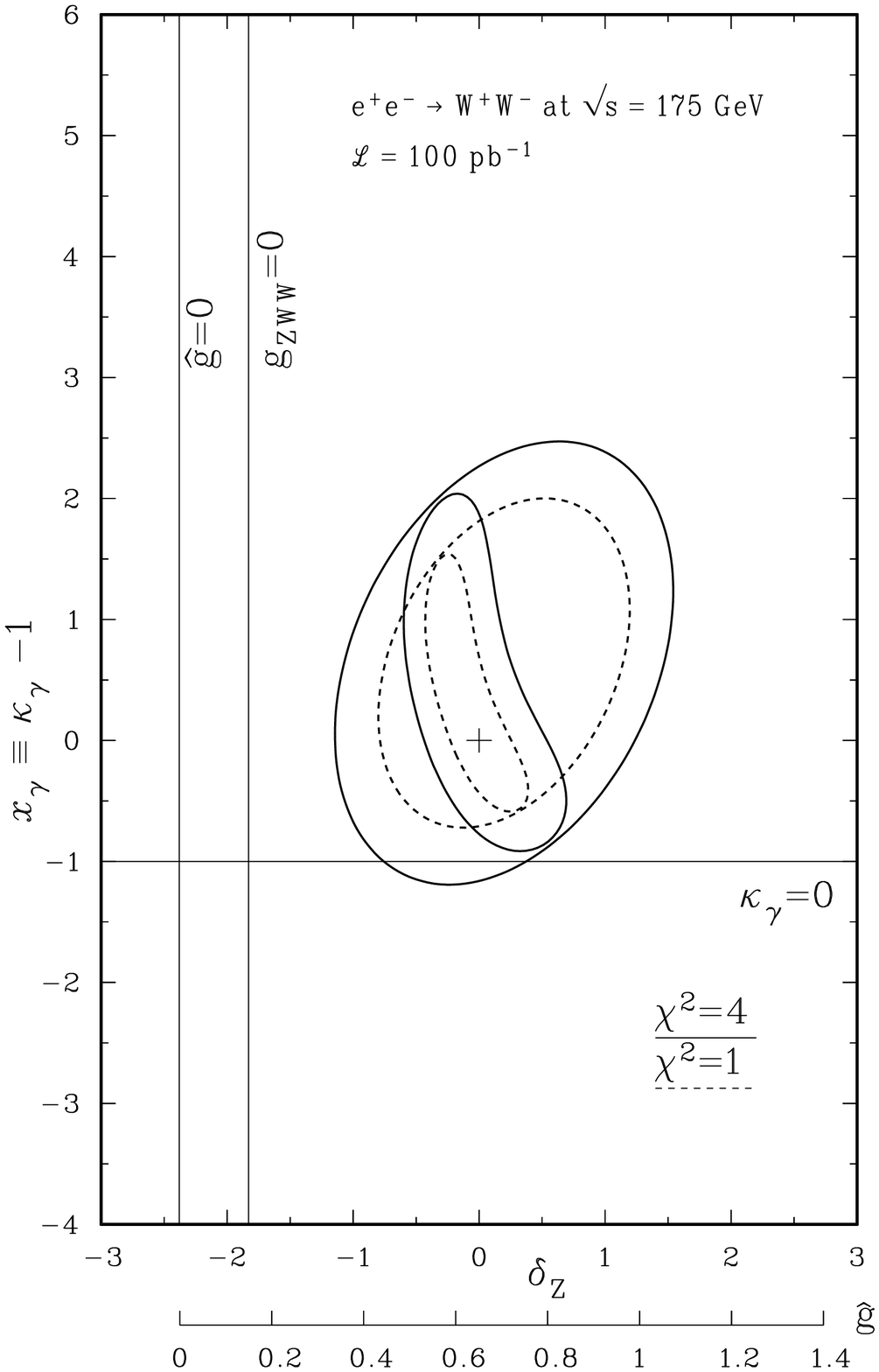,width=7.2cm,
bbllx=40,bblly=40,bburx=490,bbury=710}\cr
a&b
\end{tabular}
\end{center}
\caption{a: Detecting the existence of a non-Abelian vector-boson coupling, 
\protect{$\hat{g}\neq 0$}, at LEP~2.
b: Detecting a non-zero anomalous magnetic dipole moment,
\protect{$\kappa_\gamma\neq 0$}, of the \protect{$W^{\pm}$} at LEP~2.}
\label{anom}
\end{figure}
{}Figs. \ref{anom}a and \ref{anom}b from 
\citere{KS} are based on the assumption 
that future data on $e^+ e^-
\rightarrow W^+ W^-$ at an energy of 175 GeV will agree with SM predictions
within errors. Under this assumption, fig.~\ref{anom}a shows that an integrated luminosity
of $8 pb^{-1}$, corresponding to a few weeks of running at $175\GeV$
will be sufficient to
provide {\it direct} experimental evidence for the existence of a non-vanishing
coupling of the non-Abelian type, $\hat g \not= 0$,  
among the members of the vector-boson triplet
(at 95\% C.L.). Likewise, according to fig.~\ref{anom}b, 
an integrated luminosity of $100 pb^{-1}$,
corresponding to about seven months of running at LEP2, will provide direct
experimental evidence for a non-vanishing anomalous 
magnetic moment of the W boson
(at 95\% C.L.), $\kappa_\gamma \not= 0$.  

\section{Conclusions}

Let me conclude as follows:
\begin{itemize}
\item
The Z~data and the W-mass measurements require electroweak corrections
beyond fermion-loop contributions to the vector-boson propagators. 
\item
In the Standard Model such corrections are provided by bosonic loops. The
dominant bosonic correction, needed for agreement with the data can be traced
back to the difference in scale between $\mu$ decay, entering via $G_\mu$, and
W or Z decay. While not being sensitive to the Higgs mechanism, these
bosonic corrections depend on the non-Abelian couplings among the vector
bosons. The data accordingly ``see'' the non-Abelian structure of the Standard
Model.  
\item
The bounds on the mass, $\MH$, of the Higgs scalar are most reliably derived
from the reduced set of data containing $\swbar^2 , \MW , \Gl , \Mt^{{\rm
exp}}$ and $\alpha (\MZ^2)$ besides $\MZ$ and $G_\mu$. At 95\% C.L. the 1996 set
of data implies $\MH \lsim 550$ GeV and $\MH \lsim 800$ GeV, 
depending on whether
$\swbar^2$(LEP+SLD) or $\swbar^2$(LEP) is used as input. 
These bounds are quite remarkable, as for the first time they seem to fairly
reliably predict a Higgs mass in the perturbative region of the SM.
\item
Since the ``$\Rb$-crisis'' has meanwhile been resolved by our experimental
colleges a short time after my talk in Erice, there is now perfect overall
agreement with the predictions of the SM, even upon including hadronic $Z$
decays in the analysis. The strong coupling, $\alps (\MZ^2)$, obtainable from
the hadronic Z-decay modes, comes out consistently with the event-shape
analysis. Various speculations on ``hadrophilic'' or ``leptophobic'' bosons do
not seem to be realized in nature.    
\item
{}Forthcoming experiments at LEP2 on $e^+ e^- \rightarrow W^+ W^-$ will allow one to
find first {\it direct} experimental evidence for the existence of non-vanishing
couplings of non-Abelian type among the vector bosons. 
\item
The available data by themselves do not discriminate a MVB from the
Standard Theory based on the Higgs mechanism. The issue of mass generation will remain
open until the Higgs scalar will be found - or something else?
\end{itemize}
 
\vspace{1cm}\noindent
{\bf Acknowledgement}
 
\vspace{0.3cm}
It is a pleasure to thank Misha Bilenky, Stefan Dittmaier, Carsten 
Grosse-Knetter, 
Karol Kolodziej, Masaaki Kuroda, Ingolf Kuss and Georg Weiglein for a fruitful 
collaboration on various aspects on the theory of  
electroweak interactions.
Many thanks to Antonino Zichichi 
for providing the traditional, so much inspiring scientific atmosphere in
Erice, and last not least for magnificent hospitality.

\end{document}